\documentclass{aa}  

\usepackage{array}
\usepackage{graphicx}
\usepackage{txfonts}
\usepackage{xcolor}
\usepackage{booktabs}
\usepackage{longtable}
\usepackage[normalem]{ulem}
\usepackage{soul}
\usepackage[percent]{overpic}
\usepackage[percent]{overpic}
\newcommand{\teff}{$T_{\mathrm{eff}}$}

\definecolor{blue2}{rgb}{0.3, 0.1, 0.8}

\begin{document}

   \title{S stars and s-process in the Gaia era \thanks{Based on observations made with the Mercator Telescope, operated on the Island of La Palma by the Flemish Community, at the Spanish {\it Observatorio del Roque de los Muchachos} of the {\it Instituto de Astrofísica de Canarias}.}}

   \subtitle{I. Stellar parameters and chemical abundances in a sub-sample of S stars with new MARCS model atmospheres}

   \author{S. Shetye
          \inst{1,2}
          \and
          S. Van Eck\inst{1} \and
          A. Jorissen \inst{1} \and
          H. Van Winckel\inst{2} \and
          L. Siess\inst{1}   \and
          S. Goriely\inst{1} \and
          A. Escorza \inst{2,1} \and
          D. Karinkuzhi \inst{1,3} \and
        B. Plez \inst{4}          
          }

   \institute{ Institute of Astronomy and Astrophysics (IAA), Université libre de Bruxelles (ULB)
,
              CP  226,  Boulevard  du  Triomphe,  B-1050  Bruxelles, Belgium\\
              \email{Shreeya.Shetye@ulb.ac.be}
         \and
             Institute of Astronomy, KU Leuven, Celestijnenlaan 200D, B-3001 Leuven, Belgium
                     \and
        Department of Physics, Bangalore University, Jnana Bharathi Campus, Bangalore, India 560056
        \and
        Laboratoire Univers et Particules de Montpellier, Université de Montpellier, CNRS, 34095, Montpellier Cedex 05, France\\
             }

   \date{Received ; accepted }

\abstract{}{}{}{}{}
 
  \abstract
   {S stars are transition objects between M-type giants and carbon stars on the asymptotic giant branch (AGB). They are characterized by overabundances of s-process elements. Roughly half of them are enhanced in technetium (Tc), an s-process element with no stable isotope, while the other half lack technetium. This dichotomy arises from the fact that Tc-rich S stars are intrinsically producing s-process elements and have undergone third dredge-up (TDU) events, while Tc-poor S stars owe their s-process overabundances to a past pollution by a former AGB companion which is now an undetected white dwarf, and since the epoch of the mass transfer, technetium has totally decayed.}
   {Our aim is to analyse the abundances of S stars and gain insights into their evolutionary status and on the nucleosynthesis of heavy s-process elements taking place in their interior. In particular, the location of extrinsic and intrinsic S stars in the HR diagram will be compared with the theoretical onset of the TDU on the thermally-pulsing AGB.}
   {A sample of 19 S-type stars was analysed by combining  HERMES high-resolution spectra, accurate Gaia Data Release 2 (GDR2) parallaxes,  stellar-evolution models, and newly-designed MARCS model atmospheres for S-type stars.
   Various stellar parameters impact the atmospheric structure of S stars, not only effective temperature, gravity, metallicity and microturbulence, but also C/O and [s/Fe]. We show that photometric data alone are not sufficient to disentangle these parameters. We present a new automatic spectral-fitting method that allows
   one to constrain the range of possible atmospheric parameters.}
   {Combining the derived parameters with GDR2 parallaxes allows a joint analysis of the location of the stars in the Hertzsprung-Russell diagram and of their surface abundances. For all 19 stars, Zr and Nb abundances are derived, complemented by abundances of other s-process elements for the three Tc-rich S stars. 
   These abundances agree within the uncertainties with nucleosynthesis predictions for stars of corresponding mass, metallicity and evolutionary stage. The Tc dichotomy between extrinsic and intrinsic S stars is seen as well in the Nb abundances: intrinsic, Tc-rich S stars are Nb-poor, whereas extrinsic, Tc-poor S stars are Nb-rich. Most extrinsic S stars lie close to the tip of the red giant branch (RGB), and a few are located along the early AGB.  All appear to be the cooler analogues of barium stars. Barium stars with masses smaller than 2.5 M$_\odot$ turn into extrinsic S stars on the RGB, because only for those masses does the RGB tip extend to temperatures lower  than $\sim 4200$~K, which allows the ZrO bands distinctive of S-type stars to develop.
On the contrary, barium stars with masses in excess of $\sim 2.5$~M$_\odot$ can only turn into extrinsic S stars on the E-AGB, but those are short-lived, and thus rare.      The location of intrinsic S stars in the HR diagram is compatible with them being thermally-pulsing AGB stars. Although nucleosynthetic model predictions give a satisfactory distribution of s-process elements,  fitting at the same time the carbon and heavy s-element enrichments still remains difficult.
   Finally, the Tc-rich star V915~Aql is challenging as it points at the occurrence of TDU episodes in stars with masses as low as $M \sim 1$~M$_\odot$.}
   {}

   


   \maketitle
%

\section{Introduction}\label{Sect:Intro}

S stars are late-type giants characterized by the presence of TiO and ZrO molecular bands in their optical spectrum \citep{Merrill}. The carbon to oxygen (C/O) ratio of S stars lies in the range from 0.5 to just below unity. This indicates that they are transition objects between M-type giants (C/O~$\sim$~0.4) and carbon stars (C/O~>~1) on the asymptotic giant branch \citep[AGB;][]{IbenRenzini}. The spectra of S stars also show signatures of an overabundance of s-process elements \citep{smithlambertfeb1990}. The s-process is a slow neutron-capture process which is responsible for the production of approximately half of all the elements heavier than Fe \citep{s-process}. S-process elements are transported to the stellar surface of a thermally pulsing AGB (TP-AGB) star by a mixing process called third dredge-up (TDU). A TP-AGB star undergoes several thermal pulses which are recurrent thermal instabilities affecting the thin He-burning shell. During a pulse, He-burning ashes (mostly ${^{12}}$C) are mixed in the intershell region separating the H- and He-burning shells. Following the expansion and cooling of the intershell region, the H-burning shell extinguishes, which allows the penetration of the convective envelope in the intershell region. As a result, material from the intershell is mixed in the envelope; this is the TDU. During the TDU, the convective envelope does not only transport material from the H-He intershell zone to the stellar surface but also injects protons into the carbon-rich intershell which,
if partially mixed, trigger neutron production through the $^{12}$C(p,$\gamma)^{13}$N($\beta)^{13}$C($\alpha$,n)$^{16}$O chain of reactions.

The understanding that the overabundance of s-process elements in S stars is due to the nucleosynthesis along the TP-AGB was challenged when Tc lines were observed in some but not all S stars \citep{merrill1952,scalomiller1981}. Tc has no stable isotope and ${^{99}}$Tc, the isotope produced by the s-process, has a half life of only $2.1 \times 10^{5}$~yr. This puzzle about the evolutionary status of S stars without Tc was solved when it was realized that Tc-poor S stars belong to binary systems \citep{jorissen1988,smith1988,jorissen1993}. Hence, S stars may be classified into two different types: Tc-rich as intrinsic S stars and Tc-poor as extrinsic S stars \citep{IbenRenzini}. The intrinsic S stars are luminous and cool stars on the TP-AGB producing Tc and the other s-process elements themselves. On the contrary, extrinsic S stars owe their s-process overabundances to mass transfer from a former AGB companion which is now a white dwarf. Extrinsic S stars are the cooler analogues of barium stars \citep{jorissen1988,SophieTc,2000A&A...360..196V,2000A&AS..145...51V}.

\renewcommand{\arraystretch}{1.5}
\renewcommand{\tabcolsep}{3pt}
\begin{table*}
\caption{\label{basicdata}Basic data for the S star targets. Columns 1, 2, and 3 list different identifiers: Tycho2 \citep{tycho2}, GCVS, HD, BD, and Catalog of Galactic S Stars (CGSS) entry number. Columns 4, 5, 6, and 7 list spectral type, $V$ magnitude, $V-K$ colour, and $K$ magnitude, respectively, extracted from the SIMBAD Astronomical Database \citep{simbad}. Column 8 lists the GDR2 parallax and its error. Columns~9 and 10 compare the distances obtained by simple inversion of the parallax and by Bayesian inference \citep{2018arXiv180410121B}, the latter including the 1$\sigma$ error bar.   The reddenning $E_{B-V}$ has been obtained from the \cite{gontcharov} extinction maps. Columns 12 and 13 list the $K$-[12] and $K$-[25] from the IRAS Point Source Catalog and 2MASS $K$ magnitude \citep{2MASS}.}
\centering
\begin{tabular}{ l l r l r c l l c l l c  c l} 
\hline
TYC  &Name &  CGSS & Sp. type & $V$ & $V-K$ & $K$ &   \multicolumn{1}{c}{$\varpi\pm\sigma_{\varpi}$} & $d(1/\varpi)$ & $d$(Bayes) & $E_{B-V}$&$K-[12]$ & $K-[25]$ \\ 
  &   &   &   & & & & \multicolumn{1}{c}{(mas)} & \multicolumn{1}{c}{(pc)} & \multicolumn{1}{c}{(pc)}\\
 \hline
1048-1612-1 & V915 Aql      &  1099   & S5+/2     &  8.4 & 6.3 & 2.1 & $1.97\pm0.06$ & 507 & $500 {+18 \atop -16}$ & 0.17 & 1.0 & 1.3 \\
5419-1968-1 & NQ Pup     & 422    & S4.5/2       &  7.6 & 5.3 & 2.3 & $1.56\pm0.07$ 
& 637 & 627${+30 \atop -27}$& 0.08 & 0.8 & 0.9 \\
7786-2428-1 & UY Cen     & 816    & S6/8         &  7.1 & 6.5 & 0.6 & $1.64\pm0.14$ & 608 & 604${+62}\atop{-52}$	
 & 0.15 & 1.3 & 1.8\\
494-1067-1  & HD 189581     & 1178   & S4*2      &  8.5 & 5.6 & 2.8 & $1.48\pm0.06$ & 673 & 662${+30}\atop{-28}$	
 & 0.05 & 0.5 & 0.7\\
3372-809-1  & HD 233158     & 152    & S         &  9.3 & 6.0 & 3.3 & $1.29\pm0.05$ & 773 & 757${+34}\atop{-32}$
 & 0.08 & 0.5 & 0.5\\
2675-3119-1 & HD 191589     & 1194   & M0SIb/II  &  7.3 & 4.1 & 3.1 & $2.34\pm0.06$ & 427 & 423${+11}\atop{-11}$	
 & 0.06 & 0.4 & 0.5 \\
2683-716-1  & HD 191226     & 1192   & M1-3SIIIa &  7.3 & 5.0 & 2.3 & $1.22\pm0.03$ & 817 & 798${+23}\atop{-21}$
& 0.12 & 0.3 & 0.3 \\
2636-435-1  & V530 Lyr   & 1053   & M3-5SIII     &  7.4 & 5.1 & 2.3 & $1.99\pm0.03$ & 501 & 494${+10}\atop{-9}$
& 0.02 & 0.5 & 0.6 \\
2744-435-1  & HD 215336     & 1304   & S         &  7.8 & 4.0 & 3.8 & $1.45\pm0.05$ & 686 & 673${+24}\atop{-22}$
& 0.12 & 0.6 & 0.8 \\
5641-427-1  & HD 150922     & 937    & M2S       &  7.9 & 6.3 & 1.6 & $1.51\pm0.05$ & 659 & 648${+24}\atop{-22}$	
& 0.04 & 0.3 & 0.3 \\
5989-625-1  & HD 63733      & 411    & S3.5/3    &  7.9 & 4.4 & 3.5 & $1.26\pm0.04$ & 792 & 775${+26}\atop{-25}$
 & 0.19 &  0.7 & 1.0 \\
4376-1398-1 & BD+69$^\circ$524     & 612    & S         &  9.3 & 4.9 & 4.4 & $1.05\pm0.03$ & 951 & 925${+31}\atop{-30}$
& 0.23 & 0.7 & 0.8 \\
2256-1443-1 & BD+28$^\circ$4592    & 1334   & S2/3:     &  9.4 & 4.4 & 5.0 & $0.71\pm0.05$ & 1400 & 1340${+103}\atop{-90}$	
 & 0.08 & 0.6 & 1.5 \\
1818-1043-1 & V1135 Tau  & 87     & S3/3         &  8.9 & 5.8 & 3.1 & $1.38\pm0.07$ & 722 & 709${+42}\atop{-38}$	
& 0.07 & 0.6 & 0.6 \\
6503-1173-1 & AB Col     & 174    & M3III        &  9.5 & 7.2 & 2.7 & $1.51\pm0.03$ & 660 & 648${+15}\atop{-15}$
& 0.08 & 0.7 & 0.8 \\
5971-534-1  & -             & 302    & S1*2      &  10.9& 5.2 & 5.6 & $0.69\pm0.03$ & 1430 & 1376${+66}\atop{-60}$	
 & 0.01 &- &-\\
5399-2402-1 & BD-10$^\circ$1977    & 332    & S3*1      &  9.3 & 6.2 & 3.1 & $0.70\pm0.05$ & 1413 & 1365${+118}\atop{-100}$	
 & 0.13 & 0.8 & 0.8 \\
5404-2748-1 & FX CMa     & 350    & S5/6         &  8.6 & 5.6 & 3.0 & $0.70\pm0.05$ & 1413 & 1366${+114}\atop{-98}$
 & 0.10 & 0.5 & 0.8\\
5976-138-1  & BD-22$^\circ$1742    & 318    &  S3:*3    &  10.3& 3.8 & 6.4 & $0.52\pm0.03$ & 1920 & 1827${+142}\atop{-123}$	
 & 0.14 & - &-\\
\hline
\end{tabular}
\end{table*}
The atmospheric parameter determination  is an intricate task for S stars. The atmospheric structure of S stars is not only governed by surface gravity ($\log g$), effective temperature (\teff) and metallicity ([Fe/H]), as usual, but is also severely impacted by the C/O, and to a lesser extent, by the overall s-process element abundance ([s/Fe]). In cool (mostly intrinsic) S stars, the molecular band strengths are sensitive to both [s/Fe] and C/O \citep{piccirillo, SVE17}. Valid atmospheric  parameters, as required for a reliable abundance analysis, are thus not easy to derive. Because of the absence of models accurately accounting for the specific molecular opacities (especially ZrO) present in S-star atmospheres, the analysis of S stars traditionally relied on model atmospheres of M-type stars. These  model atmospheres are however unable to handle the combined strong influence of \teff, C/O and [s/Fe] on the atmospheric structure of S stars. A grid of MARCS model atmospheres, covering the complete parameter space of S stars, was recently published to fill this need \nocite{SVE17} (Van Eck et al. 2017; hereafter SVE17).

In the present study, we combine the  parallaxes from the Gaia Data Release 2 (GDR2; \citealp{gdr2}) and the high-resolution spectra of S stars with the MARCS model atmospheres of S stars, to derive stellar parameters for S stars. Our goal is to gain insights into the evolutionary status and nucleosynthesis taking place in these complex stellar interiors. 

The paper is organized as follows. The data overview and target selection are presented in Sect.~\ref{Observations}. The procedures developed for the parameter determination of S stars are discussed in Sect.~\ref{stellarparams}. In Sect.~\ref{abundances}, we describe our abundance analysis and the line list used to derive the s-process element abundances. In particular
we discuss the Zr and Nb abundances for all our targets, because they give insight on the s-process operation temperature. 
We also study in detail the s-process abundance profiles for the three intrinsic S stars of our sample.
Finally, in Sect.~\ref{HRDdiscuss}, we present the GDR2 Hertzsprung-Russell (HR) diagram of a restricted sample of S stars.


\section{Observations and target selection} \label{Observations}

\subsection{S stars from the  Gaia catalogues} \label{tgascondition}

\begin{table}
\caption{\label{obs} List of S stars with available TGAS solution and high-resolution spectra. For HERMES spectra, the signal-to-noise ratio (S/N) is estimated around 500~nm. H and U in the last column stands for HERMES and UVES respectively.}
\centering
\begin{tabular}{l l l l}
\hline
Name & Observation date & S/N & Instrument\\
\hline
 V915 Aql        & 2016 May 27   & 50  &H \\
 NQ Pup          & 2010 Feb 19   & 110 &H\\
 HD 189581       & 2009 July 31  & 60  &H\\
 HD 233158       & 2016 April 22 & 50  &H\\
 HD 191589       & 2011 April 10 & 80  &H\\
 HD 191226       & 2011 April 9  & 110 &H\\
 V530 Lyr        & 2011 April 6  & 100 &H\\
 HD 215336       & 2009 July 2   & 120 &H\\
 HD 150922       & 2011 April 10 & 100 &H\\
 HD 63733        &  2014 April 4 & 75  &H\\
 BD +69$^\circ$524& 2016 April 23 & 50 &H\\
 BD +28$^\circ$4592& 2010 Oct 28  & 60 &H\\
 V1135 Tau       &  2016 Nov 16  & 60  &H\\
 AB Col          & 2016 Nov 09   & 60  &H\\
 TYC 5971-534-1  & 2016 Nov 10   & 40  &H\\
 BD -10$^\circ$1977& 2016 Nov 09 & 40  &H\\
 FX CMa          & 2016 Nov 09   & 50  &H\\
 BD -22$^\circ$1742     &  2016 Nov 10             &    40 &H\\
 UY Cen          &  2016 Feb 03              & 50    &U\\
\hline
\end{tabular}
\end{table}

Among the 1347 S stars of the General Catalog of S Stars \citep[GCSS Second Edition;][]{cgss}, 321 are present in the
Gaia Data Release 1 (DR1, \citealp{gaia2}), made available when the present study started. 
If we consider only targets observable from the {\it Roque de los Muchachos Observatory} (La Palma, Canary Islands, and thus eligible for a HERMES high-resolution spectrum), that is, with $\delta > -30^{\circ}$, the intersection reduces to 124 objects (see also Sect.~\ref{Spectroscopic data}).

\begin{figure}
\includegraphics[scale=0.42]{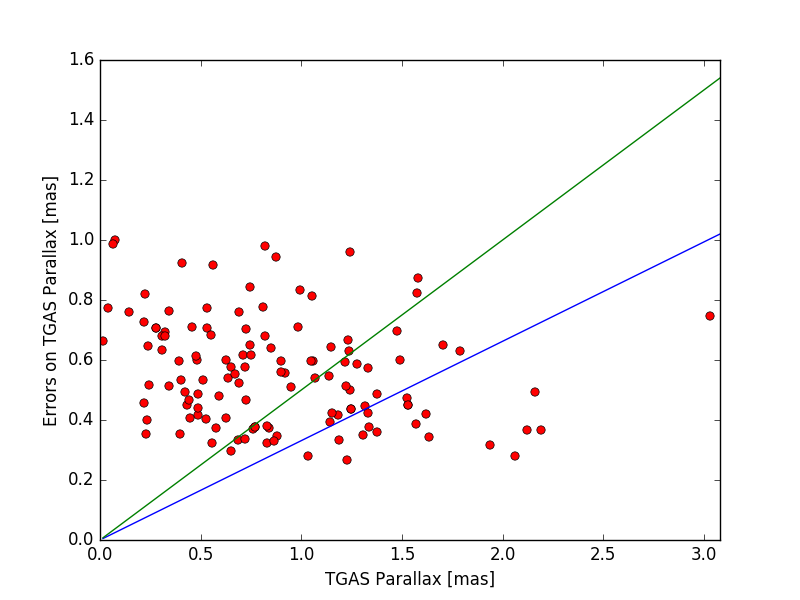}
\caption{\label{parallax} Error on the parallax $\sigma_{\varpi}$ as a function of the parallax $\varpi$ for the 124 S stars resulting from the intersection of the {\it General Catalog of Galactic S Stars} \citep{cgss} and Gaia DR1 
\citep{gaia2}, observable with the HERMES spectrograph ($\delta > -30^\circ$). The blue and green lines delineate the quality conditions defined as $\sigma_{\varpi} = 0.3 \;\varpi$ and $\sigma_{\varpi} = 0.5 \;\varpi$, respectively.}
\end{figure}

The ($\varpi$, $\sigma_{\varpi}$) plane (where $\varpi$ is the DR1 parallax and $\sigma_{\varpi}$ its error) for these 124 S stars  is presented in Fig.~\ref{parallax}.
For the present study, only stars with good-quality DR1 parallaxes are selected, that is, matching the condition {$\sigma_{\varpi} / \varpi$}~$\leq 0.3$. This condition ensures that the distance obtained by simple inversion of the parallax is not significantly biased. The positioning of individual stars on specific evolutionary tracks, as performed in Sect.~\ref{HRDdiscuss}, indeed requires 
accurate distance estimates.
These 18 S stars with Gaia DR1 parallaxes are located below the blue line in Fig.~\ref{parallax}. Our final sample of 19 S stars includes as well UY~Cen, which is not observable with HERMES but has a UVES spectrum and satisfies the condition on $\varpi/\sigma_{\varpi}$. The list of the 19 target S stars is presented in Table~\ref{basicdata}. Although the sample selection for the present investigation is, as described above, based on Gaia DR1, during the course of our study the DR2 parallaxes were released. We used these more accurate parallaxes, listed in Table~\ref{basicdata},  for the present paper. Our study was not extended to the full DR2 sample of S stars with accurate parallaxes, though, since we had not collected high-resolution spectra for all of them. The full GDR2 S-star sample analysis is therefore  deferred to a forthcoming paper.
Table~\ref{basicdata} shows that distances obtained by simple inversion of the parallaxes are compatible within 1$\sigma$ with the Bayesian estimate provided by \citet{2018arXiv180410121B}. In the remainder of the paper, we will thus use distances obtained from a simple inversion of the parallax.

In fact, many bright stars with $G \leq 7$ (where $G$ is the Gaia magnitude) and 'extremely blue and red sources' are missing from the Gaia DR1 \citep{gaia2}. The red cut introduces a bias against intrinsic S stars in our sample as intrinsic S stars are redder, on average, than extrinsic S stars: $(B-V)_{\rm ext} \approx 1.20$ as compared to $(B-V)_{\rm int} \approx 1.65$ \citep{2000A&AS..145...51V}. Therefore, many among the most evolved (intrinsic) S stars are  absent from our sample.

\subsection{High-resolution spectra} 
\label{Spectroscopic data}

The availability of HERMES\footnote{HERMES is an acronym for High Efficiency and Resolution Mercator Echelle Spectrograph.} high-resolution spectra \citep{raskin} is needed to derive accurate atmospheric parameters of S stars. The HERMES spectrograph is mounted on the 1.2m Mercator Telescope at the Roque de Los Muchachos Observatory, La Palma (Canary Islands).  It is a fibre-fed spectrograph in a temperature-controlled environment which ensures good wavelength stability. HERMES spectra have a wavelength coverage of 380~--~900~nm with a spectral resolution of $R=85 000$. The spectra are automatically reduced through a dedicated pipeline at the end of the observing night. 

The 18 S-type stars matching the conditions mentioned in Sect.~\ref{tgascondition} all have $V$ < 11, and are thus easily observable with HERMES. The observation log is given in Table~\ref{obs}. The spectrum of one additional intrinsic S star, UY Cen,
was obtained from the UVES POP database\footnote{https://www.eso.org/sci/observing/tools/uvespop/interface.html} \citep{POP}.
The TGAS parallax of UY~Cen satisfies the parallax condition (Sect.~\ref{tgascondition}) but with a declination $\delta = -44^{\circ}$, it is not observable with HERMES. UY~Cen is an interesting benchmark star already studied in SVE17.

\subsection{Classification based on Tc lines} 
\label{Sect:Tc}

\begin{figure*}[!htbp]     
\begin{centering}
    \mbox{\includegraphics[scale=0.31]{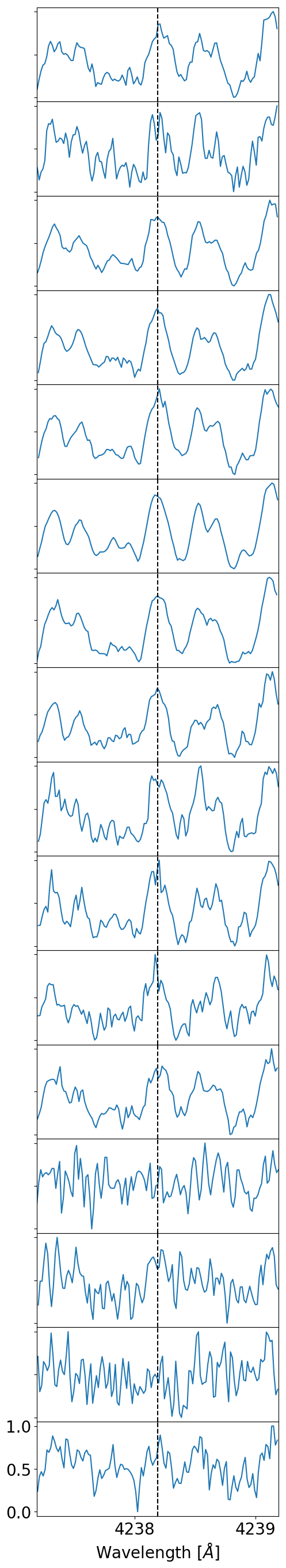}}   
    \hspace{0.1px}
    \mbox{\includegraphics[scale=0.31]{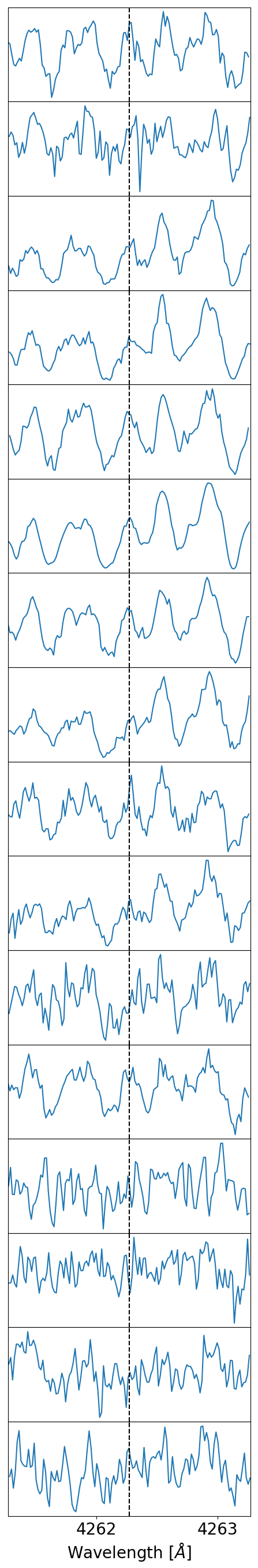}}
    \hspace{0.1px}
    \mbox{\includegraphics[scale=0.31]{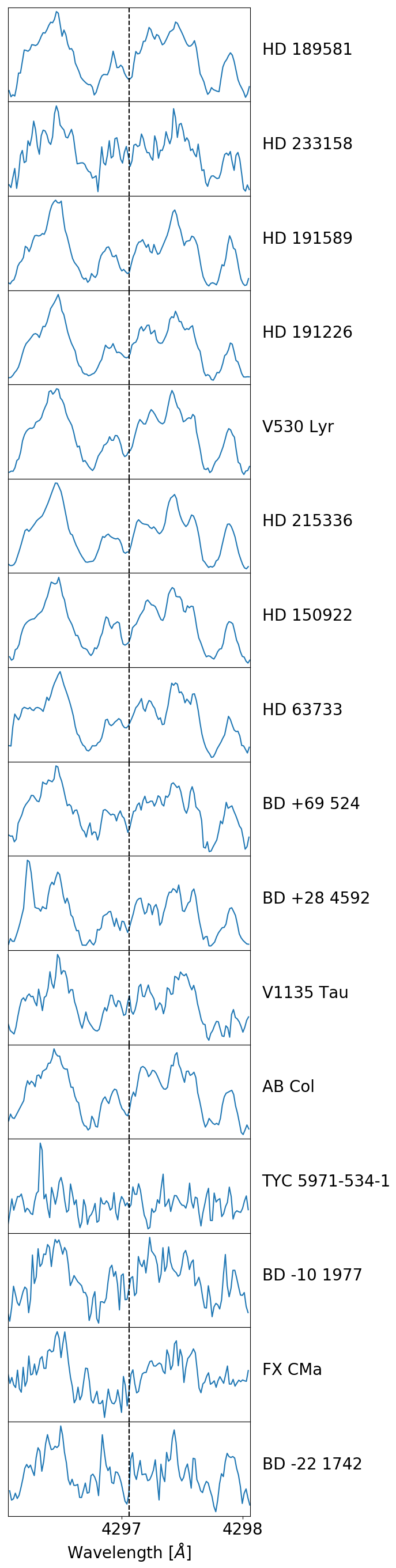}}
    \caption{\label{Tc1}
    The spectral region around the three
    (4238.19, 4262.27 and 4297.06~\AA) ultra-violet Tc I lines in Tc-poor S stars. The observed spectra have been normalized using unity-based normalization.} 
\end{centering}
\end{figure*}
\begin{figure*}[!htbp]    
\begin{centering}
    \mbox{\includegraphics[scale=0.4]{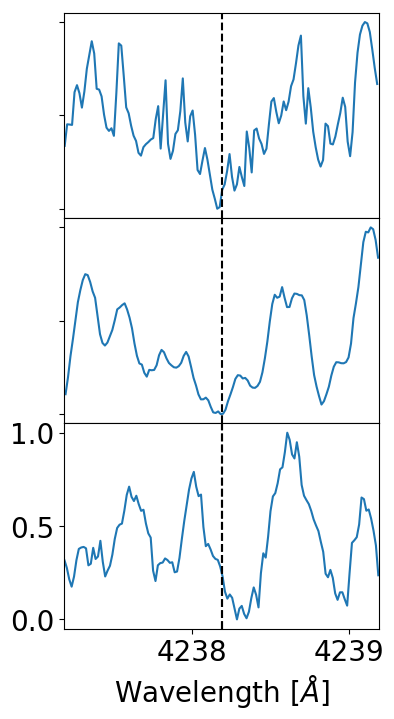}}   
    \hspace{0px}
    \mbox{\includegraphics[scale=0.4]{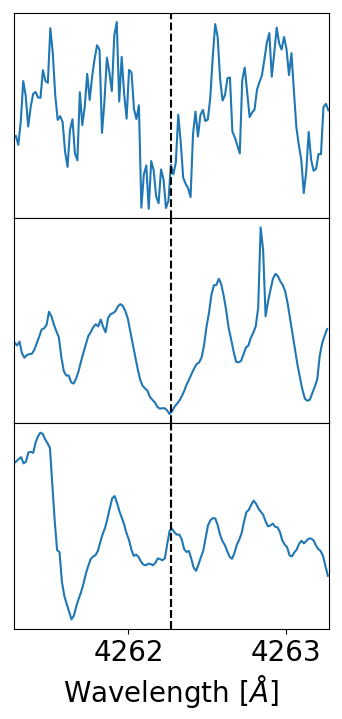}}
    \hspace{0px}
    \mbox{\includegraphics[scale=0.4]{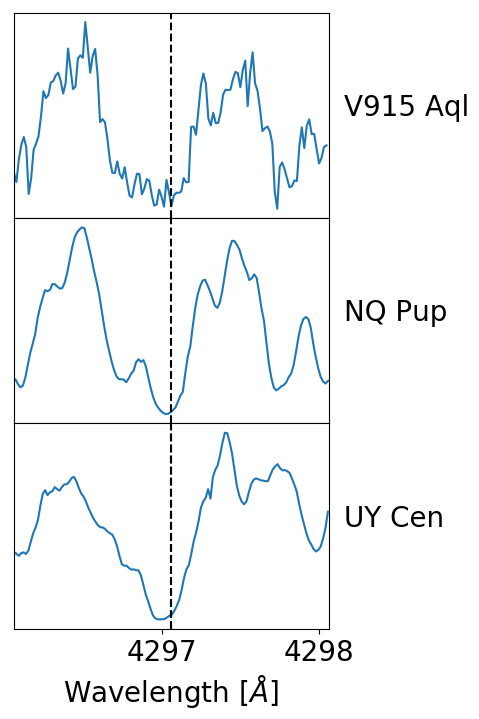}}
    \caption{\label{Tc2}Same as Fig.~\ref{Tc1} for Tc-rich S stars.}
\end{centering}
\end{figure*}

A S/N ratio larger or equal to 30 in the $V$ band is needed to detect Tc lines in S stars spectra \citep{hipp}. For this purpose, we use the three strong Tc~I resonance lines  located at $4238.19$~\AA, $4262.27$~\AA~and $4297.06$~\AA. As mentioned by \cite{Tclines} and \cite{SophieTc}, all three lines are severely blended but can nevertheless be used to distinguish intrinsic from extrinsic S stars. We classified 16 stars as Tc-poor (Fig.~\ref{Tc1}) and 3 stars as Tc-rich (Fig.~\ref{Tc2}), namely NQ~Pup, V915~Aql and UY~Cen. There is a good correlation between the diagnostics provided by the  $4238.19$~\AA\  and $4262.27$~\AA\  features. Because  the $4297.06$~\AA~Tc line is blended by other s-process element lines (especially by Nb), even when the star is devoid of Tc, the strength of that  feature varies from one Tc-poor star to another (see Fig.~\ref{Tc1}), depending on the average s-process overabundance. 

UY~Cen is an ambiguous case. It shows a clear absorption feature at $4238.19$~\AA\  and $4297.06$~\AA, thus hinting at its Tc-rich nature, but the absorption feature is not as clear at $4262.27$~\AA~(a fact also reported by \citealp{SophieTc}). Based on the two lines providing consistent results, we classify UY~Cen as Tc-rich; this classification is further supported by its infrared excess  (Sect.~\ref{IR} and Fig.~\ref{SEDs}).

The presence of Tc was first detected in V915~Aql by \cite{v915tcdet}, and Fig.~\ref{Tc2} clearly confirms the Tc-rich nature of that star. NQ~Pup was classified as intrinsic S star by \cite{smith1988}; we confirm this classification, as we do for the Tc-poor nature of HD~189581, HD~191589, HD~191226, HD~150922 and HD~215336 as previously found by \cite{smith1988} and of BD~-28$^\circ$4592 and V1135~Tau as formely published in \cite{jorissen1993}. 
\cite{wang} classified HD~233158 and BD~-10$^\circ$1977 as Tc-deficient from their location in the  $(K-[12])-([12]-[25])$ photometric plane. Finally, \cite{otto} found BD~+69$^\circ$524 to be Tc-poor from the IRAS photometry. We confirm these results from the analysis of our high-resolution spectra (Fig.~\ref{Tc1}).

For some objects however there are some disagreements with previous assignments. V530~Lyr was classified as Tc-rich by \cite{smith1988} and \cite{hipp}, but they mention that their classification for this star was uncertain due to the blend at $4262.27$~\AA. Similarly, HD~63733 was classified as Tc-rich by \cite{smith1988} and \cite{hipp}, though the latter mentioned an uncertainty on this classification.  However, from the features at $4238.19$~\AA\  and $4262.27$~\AA, we re-classify unambiguously both stars as Tc-poor.

Finally, there was no former classification available for AB~Col, TYC~5971-534-1, FX~CMa and BD~-22$^\circ$1742 in the literature. From their Tc features (Fig.~\ref{Tc1}), we have tagged all four stars as extrinsic S stars.
The final Tc-rich/poor assignment for our sample stars is listed in Table~\ref{finalparams}.

To confirm the paradigm that Tc-deficient S stars are always member of binary systems, we investigated the binary status of our extrinsic S stars. HD~63733, HD~189581, HD~191226, HD~191589, HD~215336,  V530~Lyr,   BD~+28$^\circ$4592, and V1135~Tau are known to be binary systems (\citealp{jorissen1988, brown, jorissen1993}). The radial velocity (RV) for the other extrinsic S stars was not monitored to check their binarity.  

Similarly, for the three intrinsic S stars, there is no RV monitoring available to test their binarity. However, the frequency of occurrence of binary systems among intrinsic S star is expected to be similar to that among normal (that is, non s-process-enriched) stars. For a discussion of this  topic, see for example Table 9.1 and Sect.~9.3 in \citet{Jorissen-2004} as well as \citet{jorissen1993}. We note that \cite{brown} classified NQ~Pup and V915~Aql as single stars from the absence of the \ion{He}{I} features at $10~830$~\AA, contrary to the situation prevailing for binary systems with a white dwarf companion where this feature is observed.



\section{Stellar parameters determination} \label{stellarparams}

 In an attempt to obtain the stellar parameters for S-type stars solely based on photometric data, we designed a spectral energy distribution (SED) fitting routine called S4U (S StarS SED fitting Utility). SED fitting has been used as a successful technique for atmospheric parameter determination of barium  \citep{anaBa} and post-AGB stars \citep{michel}. Hence, we tested the capability of photometric observations alone to constrain the extended parameter space of S-type stars with S4U (Sect.~\ref{Sect:S4U}). 
 Because photometry alone cannot lift the degeneracies between the S star stellar parameters, we later developed a second method (Sect.~\ref{Sect:spectralfitting}) based on comparing a high-resolution HERMES spectrum with the grid of synthetic spectra from SVE17. 
We shall confirm, as already mentioned by SVE17, that  high-resolution spectroscopy is needed for the careful disentanglement of the atmospheric parameters of S stars.

\subsection{ S4U: S StarS SED Fitting Utility} 
\label{Sect:S4U}

The S4U code  compares the available photometric observations of a given star with those computed from the synthetic spectra of the grid of S-star models (SVE17). S4U is an extension of the tool described in \cite{degroote} for the SED fitting with Kurucz atmospheric models. The 5 dimensions of the S4U parameter space are \teff, $\log g$, reddening $E_{B-V}$, [C/Fe], and [s/Fe]. The inclusion of [C/Fe] and [s/Fe], which are atmospheric parameters important for S stars (SVE17 and Sect.~\ref{Sect:Intro}), makes S4U a specific tool for SED fitting of S stars. 

\subsubsection{Method}
S4U collects first the photometry of the targets from the SIMBAD database \citep{simbad}. These photometric data are carefully cleaned, removing all the saturated data points as well as  those flagged as unreliable sources by SIMBAD. S4U cannot be used for  large-amplitude pulsators like Mira variables since the  photometric data were most likely collected at different epochs. S4U uses 2 grids of model SEDs, separately for metallicities [Fe/H] = 0.0 and -0.5. These model SEDs are medium-resolution synthetic spectra computed from the MARCS model atmospheres of the S-star grid. The parameter range covered by this grid is 2700~K -- 4000~K (with steps of 100~K) for \teff, $0 \leq \log g \leq 5 $ (with steps of 1~dex), C/O from 0.5 to 0.99 (0.50, 0.75, 0.90, 0.92, 0.95, 0.97,
0.99), [s/Fe] of 0.0, +1.0, +2.0~dex, and $[\alpha$/Fe] scaled with metallicity.
The microturbulence is fixed at 2 km/s.
Next, a photometric grid is obtained by integrating each synthetic spectrum of the model grid over the transmission curves of the various photometric filters. The photometric systems used by S4U include Hipparcos, Str{\"o}mgren, Johnson, Cousins, 2MASS, Geneva, Tycho2, Galex, DENIS, SDSS and USNO-B. The photometry has been calibrated using the Vega model of \cite{bohlin} with the zero points of \cite{maiz}. 

Only photometric bands bluer than $K$ are considered for the SED fitting, to avoid being sensitive to possible photometric excesses caused by  circumstellar dust. A $\chi^{2}$ statistics is then used to establish the goodness-of-fit for about 1 million models obtained randomly by interpolation of the photometric grid in the five atmospheric parameters listed above. The best-fitting model is of course the one with the minimum $\chi^{2}$. The error on the parameters is derived as the 1-$\sigma$ error enclosing 67\% of the model fits (see the coloured regions of Figs.~4b -- d).

\begin{figure*}
    \centering
    \begin{minipage}[t]{0.35\textwidth}
        \centering
        
        {\includegraphics[width=\textwidth]{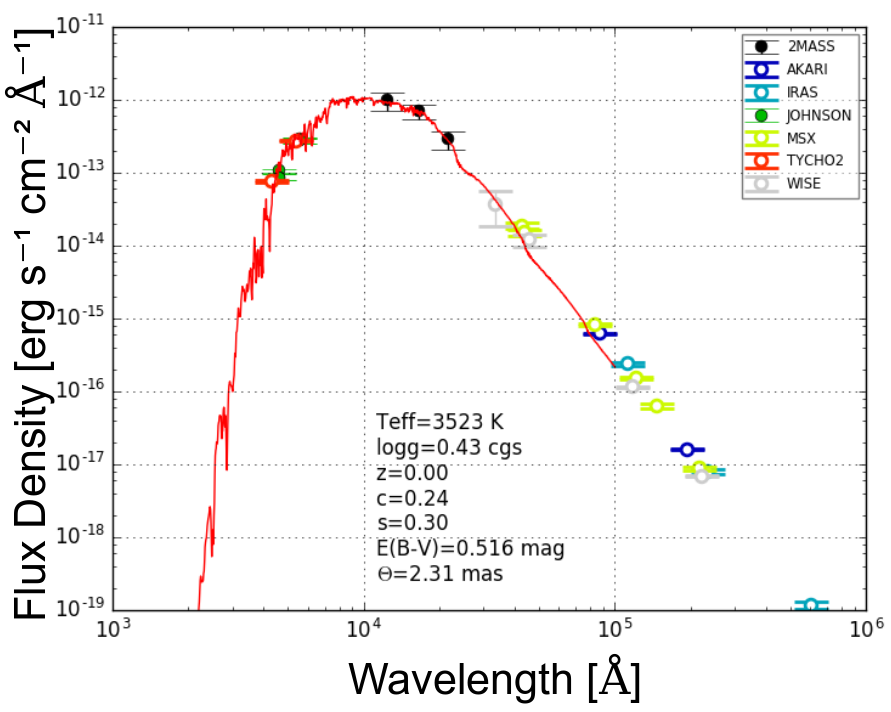}}
        \label{SED}
    \end{minipage}
    ~ 
    \begin{minipage}[t]{0.4\textwidth}
        \includegraphics[width=\textwidth]{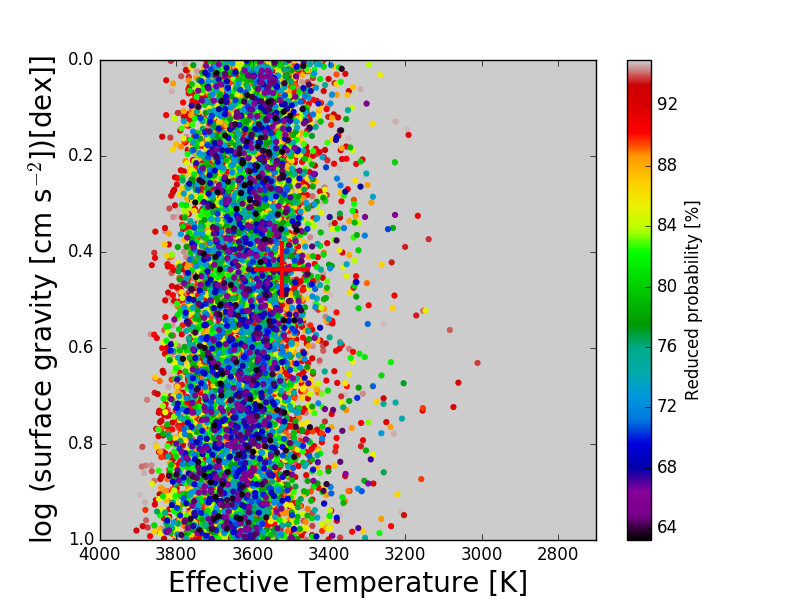}
        \label{logg}
    \end{minipage}
    \hspace{\fill}
    \begin{minipage}[t]{0.4\textwidth}
        \includegraphics[width=\textwidth]{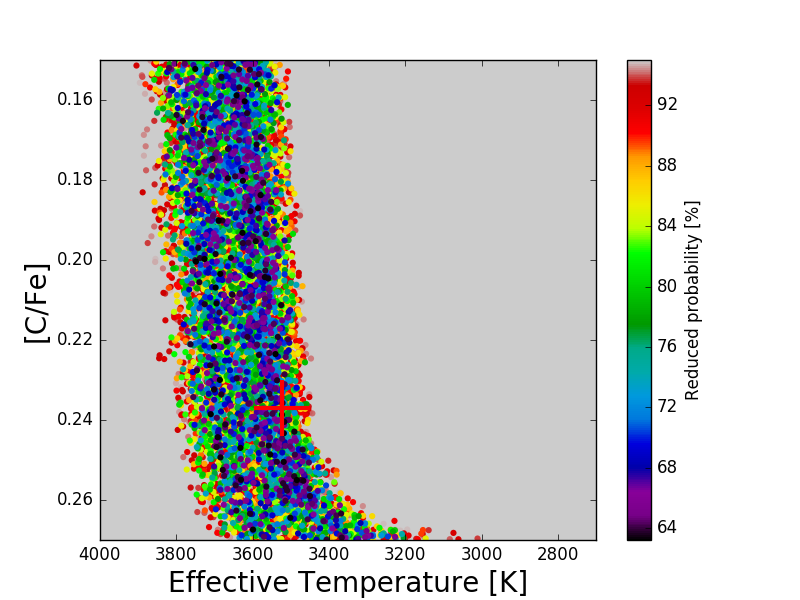}
        \label{CoverO}
    \end{minipage}
    \begin{minipage}[t]{0.4\textwidth}
        \includegraphics[width=\textwidth]{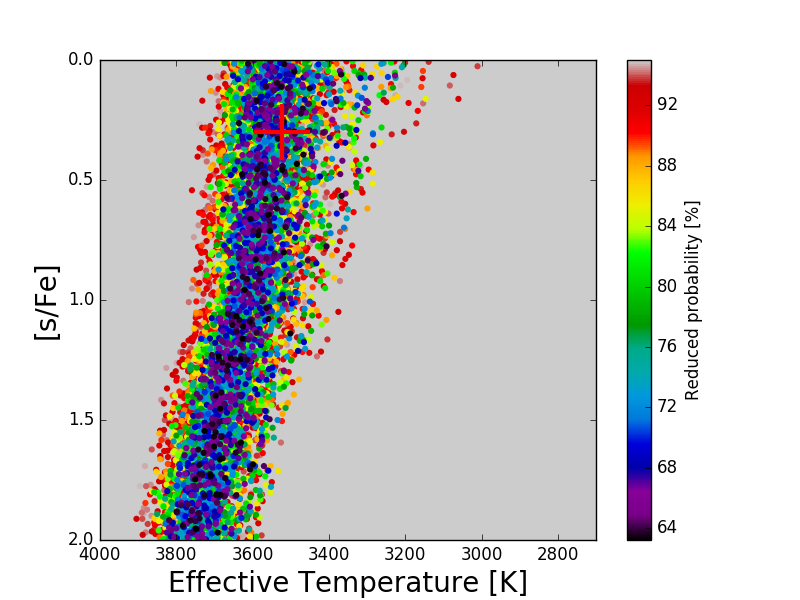}
        \label{soverFe}
    \end{minipage}
    \caption{Top left panel: the SED fitting of V915 Aql using S4U where the red line represents the MARCS model best matching the photometric observations collected from SIMBAD. Top right panel: the probability distribution plot of $\log g$ vs \teff~for V915 Aql obtained  using S4U. Bottom left panel: same as top right panel but for [C/Fe] vs \teff~and bottom right panel for [s/Fe] vs \teff. The red plus sign in top right, bottom left and right panels indicates the best  model selected by S4U.}\label{SEDresults}
\end{figure*}

\begin{table*}
\caption{\label{SEDtable} Stellar parameters for a set of Henize stars obtained from SVE17, compared with results from S4U. The values between brackets in the SVE17 columns correspond to the range covered by the models with $\chi^2_{\rm min}  \leq \chi^2 \leq 1.4 \chi^2_{\rm min}$. The values between brackets in the S4U columns correspond to the range covered by the best-fitting models within the 67\% confidence interval of the reduced probability. 
The conversion of [C/Fe] from S4U into C/O is performed using the \cite{coveroconversion} abundances. }
\centering
\begin{tabular}{c| c c c | c c c}
\hline
Star&  & SVE17  &    &   & S4U &   \\ \cline{2-7}
Name & \teff (K) & C/O & [s/Fe] & \teff (K) & C/O  & [s/Fe] \\ 
\hline
Hen 4-3  & 3500 (3500;3500) & 0.50 (0.50;0.75) & 0.00 (0.00;0.00) & 3447 (3400;3500) & 0.750 (0.50;0.899) & 0.8 (0.50;1.25) \\
Hen 4-5  & 3700 (3600;3800) & 0.50 (0.50;0.75) & 1.00 (1.00;1.00) & 3628 (3550;3650) & 0.925 (0.925;0.971) & 1.8 (1.25;2.00) \\
Hen 4-14 & 3700 (3700;3800) & 0.50 (0.50;0.75) & 1.00 (1.00;1.00) & 3705 (3630;3730) & 0.971 (0.899;0.991) & 1.8 (1.25;2.00) \\
Hen 4-16 & 3500 (3500;3600) & 0.50 (0.50;0.92) & 1.00 (0.00;1.00) & 3550 (3500;3600) & 0.751 (0.50;0.925)  & 0.9 (0.00;1.00) \\
Hen 4-19 & 3600 (3500;3700) & 0.90 (0.50;0.92) & 1.00 (1.00;2.00) & 3510 (3500;3550) & 0.991 (0.991;0.991) & 1.1 (0.75;1.25) \\
Hen 4-20 & 3500 (3500;3500) & 0.50 (0.50;0.75) & 0.00 (0.00;1.00) & 3496 (3400;3500) & 0.899 (0.899;0.925) & 1.3 (1.00;1.50) \\
Hen 4-35 & 3900 (3900;3900) & 0.50 (0.50;0.75) & 1.00 (1.00;1.00) & 3918 (3850;3950) & 0.50  (0.50;0.50)   & 1.6 (1.25;2.00) \\
Hen 4-37 & 3500 (3400;3500) & 0.50 (0.50;0.75) & 0.00 (0.00;0.00) & 3443 (3350;3500) & 0.899 (0.751;0.899) & 0.6 (0.00;1.00) \\
Hen 4-40 & 3400( 3400;3500) & 0.75 (0.75;0.92) & 0.00 (0.00;0.00) & 3465 (3450;3550) & 0.925 (0.751;0.925) & 1.2 (1.00;1.50) \\
Hen 4-44 & 3500 (3500;3500) & 0.50 (0.50;0.50) & 1.00 (1.00;1.00) & 3576 (3500;3600) & 0.899 (0.899;0.925) & 1.3 (1.00;1.50) \\
\hline
\end{tabular}
\begin{center}
\end{center}
\end{table*}

\subsubsection{Discussion}

Obtaining the stellar parameters for S-type stars from SED fitting with photometry alone turns out to be very challenging. This is due to the coupling between \teff, [C/Fe], and [s/Fe], and to the fact that  the latter two values cannot be constrained solely by broad-band photometry. 
The SED fitting (Top left panel of Fig.~\ref{SEDresults}) and the plots of reduced probability (top right, bottom left and right panels of Fig.~\ref{SEDresults}) provide a visual demonstration of the above conclusion.  

The distribution of reduced probability in the $\log g$ -- \teff~plot (top right panel of Fig.~\ref{SEDresults}) shows that S4U constrains \teff~reasonably well, since a vertical strip of dark colours (low reduced probabilities imply low $\chi^2$ values and hence best-fitting models)  is well visible. The width of this strip of dark-coloured points  shows that \teff~is constrained within 200~K for V915~Aql.
To broaden the quality check, we compared as well the S4U atmospheric parameters  with those obtained by SVE17\footnote{To be consistent in this comparison with SVE17, we computed the reddening $E_{B-V}$ using the 3D Galactic Extinction Model of \cite{drimmel} and used that value as the upper limit  on $E_{B-V}$.}
 (Table~\ref{SEDtable}).
It appears that S4U reproduces the SVE17 \teff\ within 200~K, as expected from the above-mentioned error bar. 

On the contrary, $\log g$ appears to be highly contingent upon the available  photometric bands. As explained by SVE17, good constraints on $\log g$ are provided by blue, violet, and ultraviolet filters.
The poorly constrained S4U $\log g$ value for V915~Aql is apparent from Fig.~\ref{SEDresults}, since the dark strip of well-fitting models spans the complete range on the $y$-axis.

 To avoid the degeneracy due to the sensitivity on the available photometric filters and to prevent S4U from converging to  large and unphysical  $\log g$ values of 3 -- 5  (typical of dwarf and sub-giant stars), we restricted the $\log g$ range available to S4U to 0 -- 1~dex. This is the typical range of $\log g$ obtained by SVE17 for their sample of 66 Henize S stars (Table~\ref{SEDtable}), among which no dwarf star was uncovered.

The C/O is a parameter of paramount importance for S-type star atmospheres. It is severely dependent on the adopted effective temperature and plays  a major role in fixing the TiO and ZrO band strengths in S stars. However,  the broadband photometry used by S4U is quite insensitive to these band strengths. 
This might explain why the S4U estimates of [C/Fe] and [s/Fe] are strongly discrepant with respect to those obtained by SVE17 (Table~\ref{SEDtable}). 
 Therefore, S4U cannot constrain 
 the C/O (or [C/Fe]), and this is apparent 
  in the bottom left panel of Fig.~\ref{SEDresults}. 
  The situation is similar for [s/Fe], as shown by the bottom right panel of Fig.~\ref{SEDresults} in the case of V915~Aql. This holds true for other S stars as well (Table~\ref{SEDtable}).


To conclude, a SED fitting tool like S4U based solely on broadband photometry delivers a quick initial estimate of  \teff, but  is unable to reliably constrain $\log g$, [C/Fe], and [s/Fe]. Nevertheless, the interest of this photometric attempt at determining atmospheric parameters of S-type stars is to underline the difficulties specific to this class of peculiar red giants. They call for a more sophisticated method involving high-resolution spectroscopy, as described in the next section.

\subsection{Atmospheric parameters derived from spectral fitting} 
\label{Sect:spectralfitting}

Realizing the limitations of S4U, we devised a spectral-fitting method\footnote{ 
The spectroscopic method applicable for hotter stars, using excitation and ionisation equilibrium to constrain $T_{\rm eff}$ and $\log g$, cannot be used here. Indeed, given the heavily blended and depressed spectra of S stars and the current accuracy of transition probabilities and wavelengths in molecular (mainly TiO, ZrO, LaO, VO and CH) line lists, a $\chi^2$ minimization between synthetic and observed spectra has to be performed instead.} that is more efficient at disentangling the atmospheric parameters of S stars. The method is very similar to that described in \cite{pieter2015} and SVE17, and searches for the  best-fitting MARCS model by comparing the associated synthetic spectra with  chunks of the observed spectrum, along with a couple of photometric indices. 

Since the observed high-resolution spectra are not flux-calibrated, the slope of the spectrum is not reliable over extended wavelength regions ($\ge 100$~\AA). It is therefore necessary to compare the synthetic and observed spectra over small spectral domains (about 100~\AA) that have been normalized by matching the points of maximum flux, and this is done separately within each considered chunk.

The high-resolution synthetic spectra are generated by the Turbospectrum radiative-transfer code (\citealp{rodrigo}, \citealp{turbospectrum}) using the MARCS model atmospheres from SVE17. The synthetic spectra have been convolved with a Gaussian kernel to a resolution matching that of the observed spectrum and reddened to match the reddening of the target star, as derived from the \cite{gontcharov} table with the distance computed from the DR2 parallax. The comparison between observed and synthetic spectra is then performed through  $\chi^2$-minimization, summing over all spectral pixels, and using the flux itself as an estimate of the variance on the observed spectrum. In this process, it is important to remove spectral regions contaminated by telluric lines, as well as strong lines like the Na~D doublet, whose core flux is close to zero in the observed spectrum. Since the core of the line is not well reproduced in the synthetic spectrum, it would lead to very large spurious contributions to the $\chi^2$ value if it were not discarded. 

The dereddened $V-K$ and $J-K$ photometric indices have been used as well in the comparison. They have been attributed arbitrary weights corresponding to 20000 pixels, so that they contribute to the 
$\chi^2$ in proportion to their band width. The adopted 1$\sigma$ error were $0.04$ and $0.01$ for $V-K$ and $J-K$, respectively.  Values of $\chi^2$ for every spectral chunk and photometric indices are then calculated. 

The model with the lowest total $\chi^2$ value is then selected as the best-fitting model, providing a first estimate of atmospheric parameters. Finally, we validate the stellar parameters by checking high-resolution spectral windows (for example the 50~\AA\  spectral window including the two $\lambda 7819$~\AA~and $\lambda 7849$~\AA~ Zr lines; see Figs.~ \ref{fitexample} and \ref{fitexample3} below).

\subsection{Constraining $\log g$ with DR2 parallaxes}\label{loggloops}

\begin{figure}
\includegraphics[width=9cm, height=4cm]{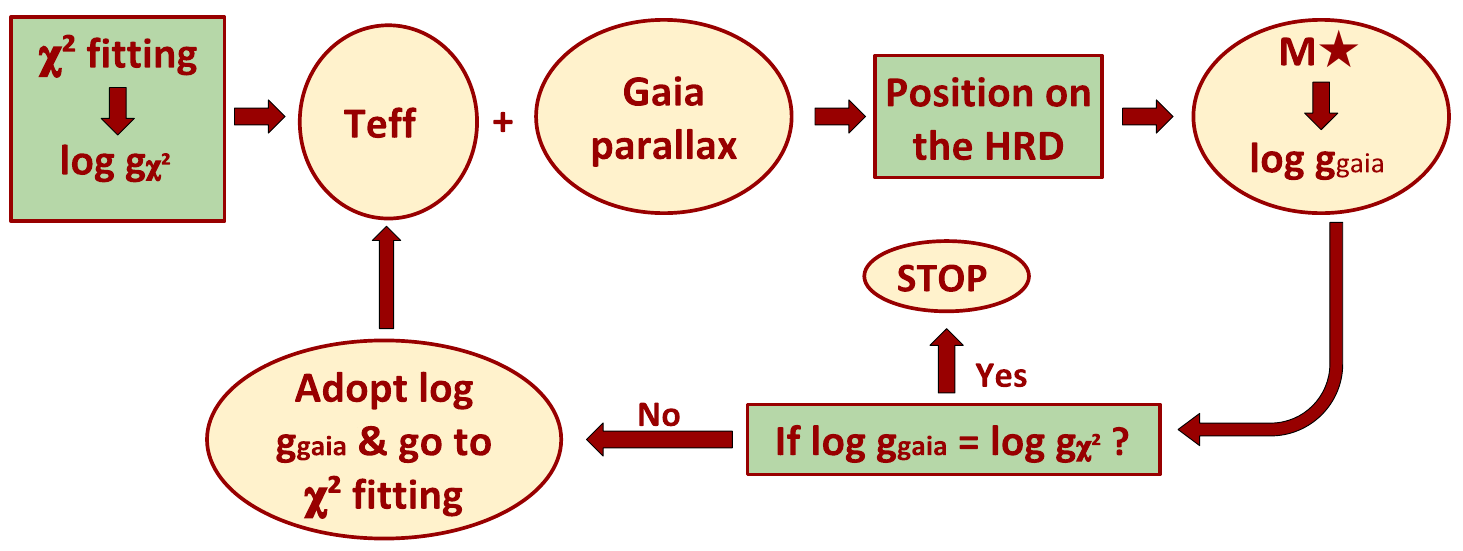}
\caption{\label{loggiter} Iterating scheme to constrain $\log g$.}
\end{figure}

To better constrain the gravities derived from the spectral-fitting method, spectral chunks in the $U$ band would be required (see the discussion in SVE17). However, these red stars have weak signals in that spectral range. Here we use instead the star location in the HR diagram (see Sect.~\ref{HRDdiscuss}) to constrain its mass and radius, and hence its gravity that we name "Gaia $\log g$" in the following. 
This Gaia $\log g$ is derived  from the relation: 
\begin{eqnarray}
\log \frac{g}{g_\odot} &=&
\log\frac{M}{M_\odot} + 4 
\log \frac{T_{\rm eff}}{T_{{\rm eff},\odot}} - \log\frac{L}{L_\odot} 
\\
&=&
\log\frac{M}{M_\odot} 
+ 4 \log \frac{T_{\rm eff}}{T_{{\rm eff},\odot}} + 0.4 \; (K + BC_K - A_K \nonumber \\
& & - 10 + 5 \log\varpi - M_{{\rm bol},\odot}) ,
\end{eqnarray}
 where $\varpi$ is the Gaia DR2 parallax expressed in mas; $A_K$ is the extinction in the $K$ band, derived as described in Sect.~\ref{Sect:spectralfitting}; $T_{\rm eff}$ and $BC_K$, the bolometric correction in the $K$ band,  are taken from the MARCS model with the lowest $\chi^2$. The mass is determined by selecting the evolutionary track of the corresponding metallicity passing closest to the star location in the HR diagram (details about the stellar models are provided in Sect.~\ref{sect:starevol}). Since the Gaia $\log g$ obtained in this way generally does not match the one from the spectral-fitting tool,  the process was iterated as illustrated in Fig.~\ref{loggiter}.

In the next iteration, we thus adopted the new Gaia $\log g$ and chose the first model in the list of best-fitting models having $\log g$ matching the Gaia $\log g$ value. 
With this new model (which could have a \teff\  and/or metallicity which differ from the previous one), we obtain a new position in the HR diagram. This iterative process is pursued until it converges towards a consistent $\log g$ value.

\subsection{Uncertainties on the atmospheric parameters}
\label{Sect:atmosph_uncertainties}

As explained above, finding the 
atmospheric parameters (\teff, $\log g$, [Fe/H], C/O and to a lesser extent [s/Fe]) for S stars  is an intricate process, and so is the estimate of their uncertainties. The method we followed is inspired by that of \citet{cayrel}. It consists in finding alternate atmospheric parameters which lead to an acceptable fit of the high-resolution spectrum. We illustrate the procedure with V915~Aql. Its best parameters are \teff\ = 3400~K, $\log g = 0$, [Fe/H] = -0.5, C/O = 0.75, and [s/Fe] = 0~dex (Table~\ref{finalparams}). In the following, this set of parameters will be called 'model~A'.  In the list of best models provided by the $\chi^2$ method (see Table ~\ref{abundtableerr3}), one of the models coming after A has \teff\ = 3500~K, $\log g = 1$, [Fe/H] = 0.0, C/O = 0.5, and [s/Fe]~=~1~dex. 
We denote this as 'model~H'. Fig.~\ref{Fig: normalandE} reveals that the synthetic spectrum for model~H fits the \ion{Zr}{I} 7819~\AA~line almost equally well as the spectral fit for model~A, though model A and model H differ from each other by one grid step for each of their parameters. We therefore consider that one grid step in each of the considered parameters represents a fair estimate of their uncertainty (and consequently, that an interpolation in the model grid would not provide any accuracy improvement). An estimate of the impact of these model-parameter uncertainties on the s-process abundance uncertainty will be discussed in Sect.~\ref{uncertainties}.

\begin{table*}
\caption{\label{finalparams} Atmospheric parameters for S stars after performing iterations on $\log g$. In the \teff\ and $\log g$ columns, the numbers between brackets denote the values spanned during the $\log g$ iterations, while in the C/O column they indicate the error on C/O from the CH regions and in the $L$ column, they indicate the luminosity error due to the GDR2 error on the parallax. The metallicities [Fe/H] have been obtained from the  Fe abundance analyses. The numbers in  brackets in the [Fe/H] column indicate the number of lines used to derive [Fe/H] and the next column indicates the standard deviation (derived from the line-to-line scatter) on [Fe/H]. $BC_K$ is the bolometric correction in the $K$ band as computed from the MARCS model atmospheres. The masses have been derived from the locations of the stars in the HR diagram compared to the STAREVOL tracks (Fig. \ref{HRD}). The last column lists whether the star is Tc-rich ("Y") or Tc-poor ("N") from the analysis of its Tc lines (Sect.~\ref{Sect:Tc}). }
\centering %
\setlength\extrarowheight{-4pt}
\begin{tabular}{l c c c l l c c l c l}
\hline \\
 Name & $T_{\rm eff}$ & $L$ & $\log g$ & [Fe/H] &$\sigma_{\rm [Fe/H]}$& C/O & [s/Fe] & $BC_K$ & Mass & Tc \\
  & (K) & ($L_{\odot}$) & & & & & &  &($M_{\odot}$) & \\
\hline \\
V915 Aql       & 3400  & 1958  & 0  & -0.5 (9)  & 0.15 & 0.75 & 0   & 3.00 &1 & Y \\
 & (3400; 3400) & (1832; 2098) & (0; 1) & & & (0.65; 0.75) & (0; 1) & & & \\
NQ Pup         & 3700  & 3059  & 1  & -0.3 (10) & 0.05 & 0.50   & 1   & 2.77 & 2.5 &Y \\
 & (3500; 3700) & (2797; 3360) & (1; 2) & & & (0.5; 0.75) & (1; 1) & & & \\
UY Cen         & 3300  & 10505 & 0  & -0.3 (7)   & 0.15 & 0.999 & 1   & 3.05 &3& Y \\
 & (3000; 3400) & (8826; 12712) & (0; 3) & & &(0.97; 0.999) & (1; 2) & & & \\
HD 189581      & 3500  & 1845  & 1  & 0.0 (12)  & 0.13 & 0.50   & 0  & 2.93&2 & N\\
 & (3500;3500) & (1692; 2019) & (1; 2) & & & (0.50; 0.75) &(0; 0)& & & \\
HD 233158      & 3600  & 1692  & 1  & -0.4 (12) & 0.16 & 0.50  & 1  & 2.84 & 1&N \\
 & (3600; 3600) & (1553; 1851) & (1; 1) & & & (0.50; 0.75) & (1; 1)& & & \\
HD 191589      & 3700 & 636  & 1 & -0.3 (12) & 0.10 & 0.75  & 1   & 2.77 &1 & N\\
 & (3700; 3800) &(604; 670)&(1; 2) & & &(0.75;0.75)  &(1; 1) & & &\\
HD 191226      & 3600  & 4767 & 1  & -0.1(12)  & 0.13 & 0.75   & 1  & 2.85 &3.5& N\\
 & (3600; 3600) &(4511; 5046) & (0; 1) & & & (0.50; 0.90)&(1; 1) & & & \\
V530 Lyr       & 3500 & 1550  & 1  & 0.0 (12)  & 0.10 & 0.50   & 1  & 2.93 &1.5& N\\
 & (3500; 3600)& (1493; 1610)&(1; 3)& & &(0.50; 0.75)&(1; 1)& & & \\
HD 215336      & 3700  & 913  & 1  & 0.0 (10)  & 0.12 & 0.50 & 1   & 2.77 &1.5& N\\
& (3700; 3700)&(854; 979)&(1; 1)& & & (0.50; 0.75)&(1; 1)& & & \\
HD 150922      & 3600  & 5614  & 0  & -0.5 (11) & 0.12 & 0.50 & 1  & 2.86 &2.5& N\\
 & (3600; 3600) & (5234; 6036) & (0; 1) & & & (0.50; 0.75) & (1; 1) & & &\\
HD 63733       & 3700 & 1614  & 1  & -0.1 (13) & 0.13 & 0.50 & 1  & 2.78 &2.5& N\\
 & (3700; 3700) & (1510; 1728) & (1; 2) & & & (0.50; 0.75) & (1; 1)& &  &\\
BD +69$^\circ$524     & 3600  & 955  & 1  & -0.4 (13) & 0.11 & 0.50 & 0   & 2.84 &1& N\\
 & (3600; 3600) & (893; 1023) &(1; 2) & &  &(0.50; 0.75)& (0; 0) & & & \\
BD +28$^\circ$4592    & 3700  & 1196   & 1  & -0.1 (15) & 0.12 & 0.75 & 1   & 2.77 &2& N\\
 & (3700; 3800) & (1036; 1396) & (1; 2) & & & (0.50; 0.90)&(1; 1) & & & \\
V1135 Tau      & 3400  & 1601 & 1  & -0.2 (14) & 0.14 & 0.50 & 1   & 2.94 &1& N \\
 & (3400; 3500) & (1435; 1798) & (0; 1) & & & (0.5; 0.95) & (1; 1) & & & \\
AB Col         & 3500  & 1940  & 1 & 0.0 (17)  & 0.14 & 0.50  & 1   & 2.92 & 2&N\\
 & (3300; 3500) & (1852; 2036) & (1; 2) & & & (0.50; 0.75) & (1; 1) & & &\\
TYC 5971-534-1 & 3600  & 666   & 1  & -0.1 (15)  & 0.18 & 0.90 & 1   & 2.83 & 1&N\\
 & (3600; 3600) & (608; 734) & (0; 2) & & &(0.50; 0.90) & (1; 2) & & & \\
BD -10$^\circ$1977    & 3500  & 6016  & 0  & -0.5(16)   & 0.15 & 0.50 & 1   & 2.94 &3& N \\
 & (3500; 3600) & (5153; 7117) & (0; 1) & & & (0.50; 0.90) & (1; 1) & & &\\
FX CMa         & 3500  & 7032  & 1  & 0.0 (8)   & 0.14 & 0.97& 1 & 2.91 &4& N \\
 & (3500; 3500) & (6021; 8319) & (1; 1) & & & (0.95; 0.97) &(1; 1)  & & & \\ 
BD-22$^\circ$1742     & 4000  & 775  & 1  & -0.3 (11) & 0.09 & 0.75 & 0  & 2.54 & 2&N \\
 & (4000; 4000) & (671; 905) & (1; 5) & & & (0.50; 0.90)&(0; 1) & & & \\
\hline
\end{tabular}
\end{table*}


\section{Abundance determination} \label{abundances}

\begin{figure*}
\centering
\includegraphics[scale=0.4]{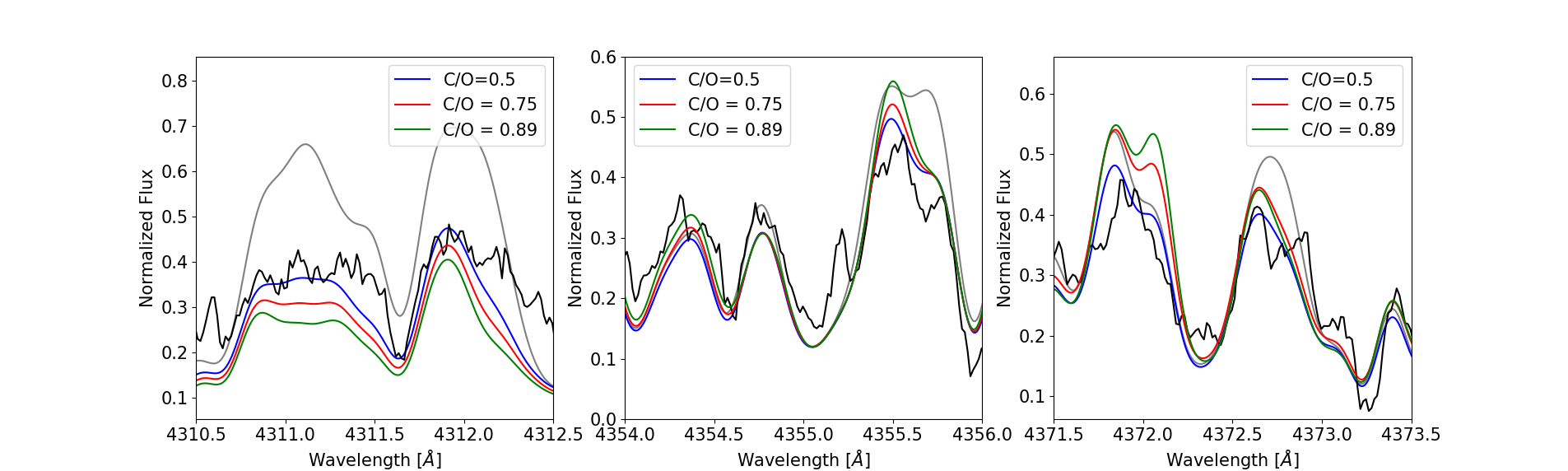}
\caption{\label{fitexampleCoverO}
 Carbon abundance derived from  a comparison between the observed spectrum of the extrinsic S star HD~189581 in the CH G-band (black line) and the synthetic spectrum for different C/O values. The best agreement is obtained for C/O$=0.5$. The grey line presents the synthesis without the CH linelist. }
\end{figure*}

\begin{figure*}
\centering
\includegraphics[scale=0.4]{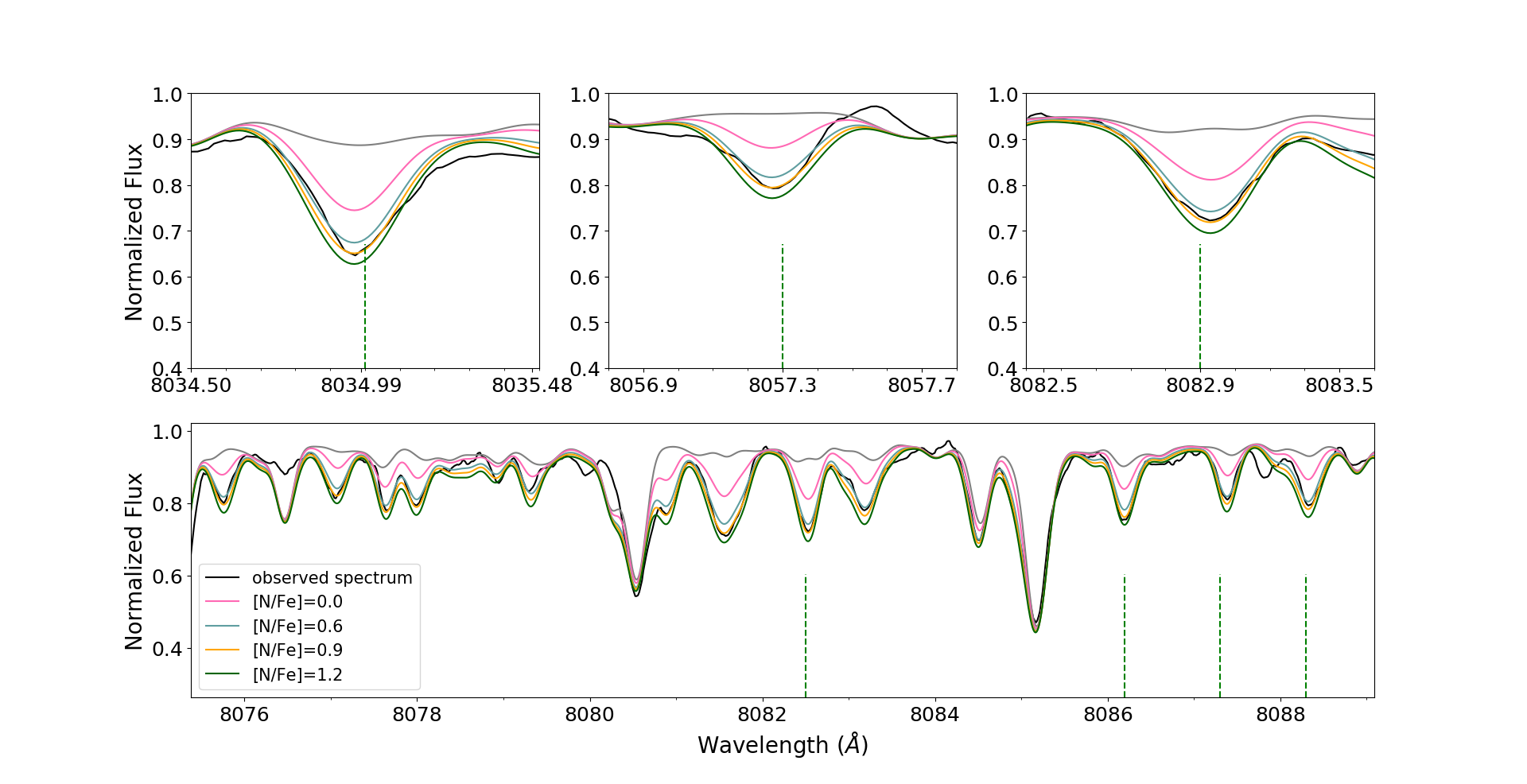}
\caption{\label{fitexampleCN}Bottom panel: comparison  between the synthetic and observed spectra for V530~Lyr in the 8076 - 8088 \AA~range containing CN lines. The top three panels present a $\pm~1$~\AA~zoom around the CN lines at 8035~\AA, 8057.3~\AA, and 8082.9~\AA\ (from left to right). The green dashed lines mark the CN lines used by \cite{tibaulth} in barium stars. The grey line presents the synthesis without the CN line list. The value [N/Fe] = 0.9 has been adopted.}
\end{figure*}

\begin{figure*}
\centering
\includegraphics[scale=0.35,trim={1cm 1cm 0.5cm 1.1cm}]{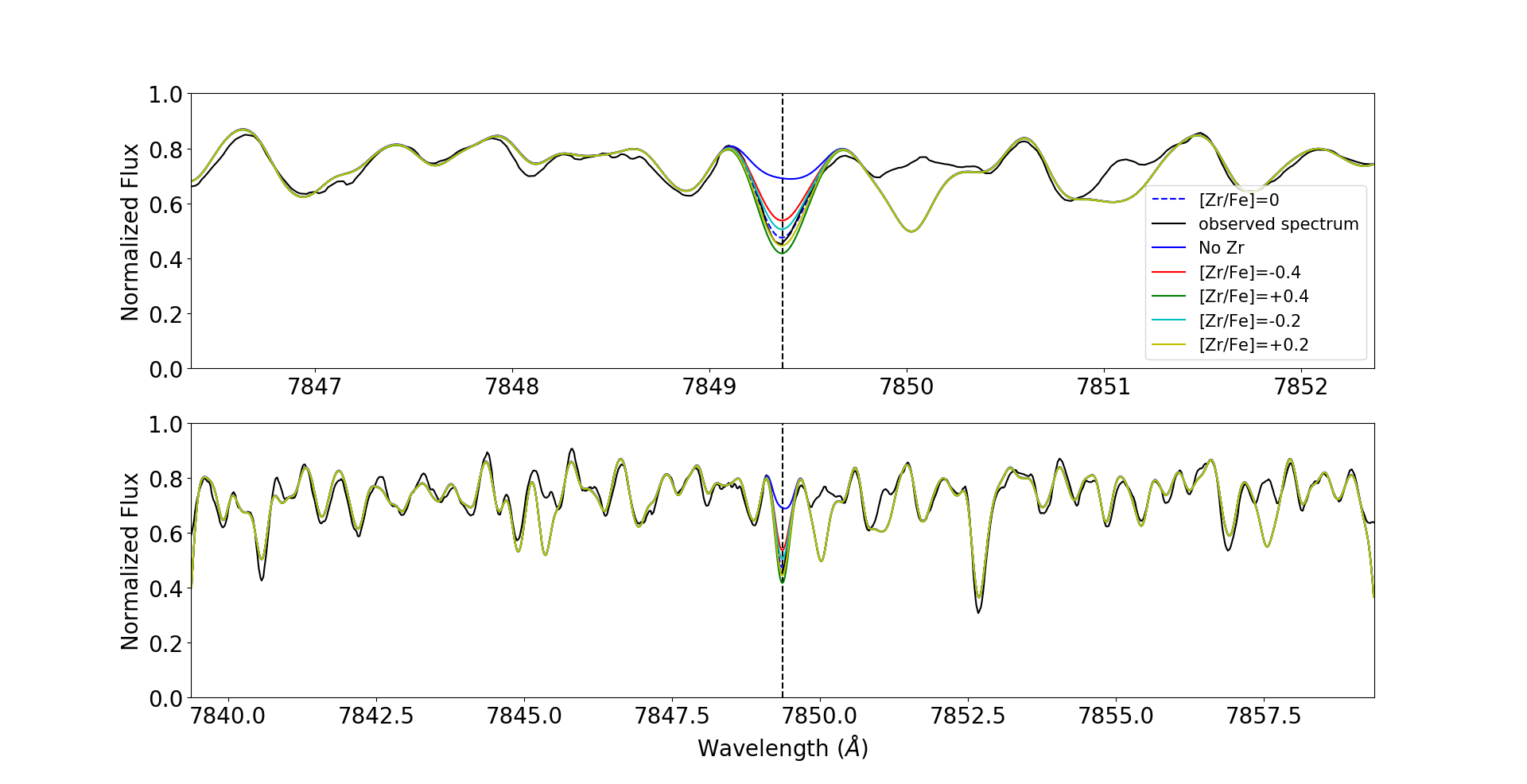}
\caption{\label{fitexample} Illustration of the quality of the match between observed and synthetic spectra obtained for the extrinsic S star V530~Lyr around the Zr line at 7849.37~\AA. The upper panel presents a $\pm~3$~\AA{} zoom.}
\end{figure*}

\begin{figure*}
\centering
\includegraphics[scale=0.35,trim={1cm 1cm 0.5cm 1.1cm}]{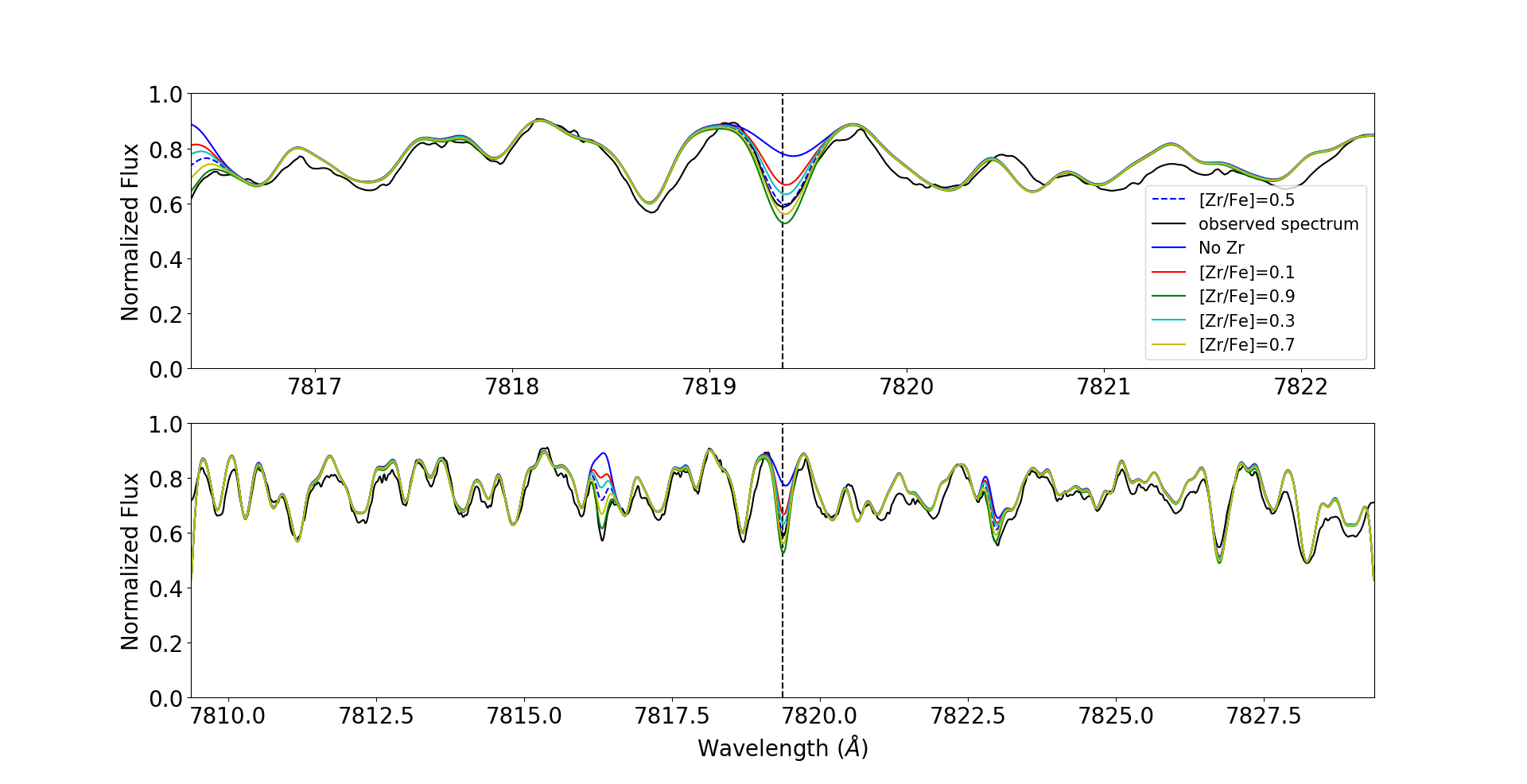}
\caption{\label{fitexample3}Same as Fig.~\ref{fitexample} for the intrinsic S star V915~Aql around the Zr line at 7819.37~\AA.}
\end{figure*}

The abundances for all the considered elements were derived by comparing the observed spectra with synthetic spectra generated by Turbospectrum using MARCS model atmospheres. We have used the same molecular and atomic line lists as SVE17.

\subsection{C, N, O}
The spectra of S stars are dominated by oxygen-bearing molecular bands. An accurate estimate of the C, N, and O abundances is needed as a prior to derive accurate atomic abundances. The carbon abundance is determined from sensitive features in the CH band around 4300~\AA~(Fig.~\ref{fitexampleCoverO}). Because the $\lambda$~6300.3~\AA\  [\ion{O}{I}] line lies in a severely blended region, it is not possible to use it for deriving the O abundance. Therefore, we use instead a metallicity-scaled solar O abundance value throughout our study. The C/O was varied over the whole range 0.50 -- 0.99 and synthetic spectra with those values were compared to the observed spectrum in order to confirm that the spectral-fitting routine indeed provides the most appropriate C/O.
Since we derived the carbon abundance from the CH G band and the C/O from an overall spectral fit, we may combine these values to derive the oxygen abundance. We checked that it was indeed consistent with the assumed metallicity-scaled solar O abundance. 
 We found an agreement better than 0.2 dex in all our stars, except for TYC 5971-534-1,
for which the discrepancy in $\log \epsilon_O$ amounts to 0.26 dex. 

 The uncertainty on C/O was estimated from the values of the C/O providing an acceptable fit to the CH G-band.
 
The nitrogen abundance was then derived from the CN lines available in the 7900~--~8100~\AA\  range (Fig.~\ref{fitexampleCN}), using in particular the lines listed in \cite{tibaulth}. We note, however, that for stars with C/O~$\ga 0.9$, the CN lines are strong and their sensitivity to the N abundance is therefore limited, making the N abundance uncertain. 

\subsection{Metallicity [Fe/H] }
\label{Sect:metallicity}

Metallicity is derived from Fe line synthesis (in the range $ \lambda 7300 - \lambda 8700$~\AA).  Metallicities along with the number of  lines used and their standard deviations are listed in Table~\ref{finalparams}; the full line list is given in Sect.~A.1. 

\subsection{s-process elements}

\begin{table*}
\caption{\label{ZrNbtable} C, N, Zr, and Nb abundances in extrinsic S stars. }
\centering
\begin{tabular}{l r c c c c c}
\hline
 Name & [C/Fe] & \multicolumn{1}{c}{[N/Fe]} & [Zr/Fe] & $\sigma_{\log \;\epsilon_{Zr}}$ & [Nb/Fe] & $\sigma_{\log\; \epsilon_{Nb}}$ \\
\hline
HD 189581      & -0.03 & 0.6  & -0.1 & 0.14 & 0.0 & 0.07 \\
HD 233158      & 0.17  & 0.7  & 0.3  & 0.14 & 0.3 & 0.17 \\
HD 191589      & 0.15  & 0.5  & 0.4  & 0.14 & 0.7 & 0.05 \\
HD 191226      & 0.15  & 0.7  & 0.4  & -    & 0.5 & 0.21 \\
V530 Lyr       & -0.03 & 0.9  & 0.2  & 0.07 & 0.3 & 0.27 \\
HD 215336      & -0.03 & 1.1  & 0.5  & 0.07 & 0.7 & 0.19 \\
HD 150922      & 0.17  & 0.4  & 0.4  & 0.21 & 0.2 & - \\
HD 63733       & -0.03 & 1.4  & 1.0  & -    & 1.2 & 0.17 \\
BD +69$^\circ$524     & 0.17  & 0.6  & 0.2  & 0.14 & 0.4 & - \\
BD +28$^\circ$4592    & 0.15  & 0.5  & 0.9  & 0.07 & 1.3 & 0.17 \\
V1135 Tau      & 0.15  & 0.2  & 0.4  & -    & 0.5 & 0.12 \\
AB Col         & -0.03 & 0.2 & 0.0  & 0.14 & -0.1& 0.20 \\
TYC 5971-534-1 & 0.28  & 0.07  & 0.2  & 0.14 & 0.5 & 0.08 \\
BD -10$^\circ$1977    & 0.13  & 0.6  & 0.07 & 0.21 & 0.19 & 0.10 \\
FX CMa         & 0.26  & 1.6  & 1.2  & -    & 1.4 & 0.17 \\
BD -22$^\circ$1742    & 0.35  &-0.1  & 0.2  & 0.28 & 0.4 & - \\
\hline
\end{tabular}
\end{table*}

Abundances of Zr and Nb are derived for all our programme stars (Tables~\ref{ZrNbtable} and~\ref{abundtable}). The Zr abundances were measured from the two \ion{Zr}{I} lines with transition probabilities from laboratory measurements \citep{Zrlines} 
at 7819.37~\AA~and 7849.36~\AA\ (see Figs.~\ref{fitexample} and \ref{fitexample3}). 
As discussed by \citet{pieter2015} and \citet{drisya}, the above  \ion{Zr}{I}
lines  systematically  lead  to \ion{Zr}{I} abundances 0.30~dex lower than 
abundances derived from lines  with  transition  probabilities  from  \citet{CB}. We favour the former lines as they rely on measured transition probabilities. 

The Nb abundances for our stars were estimated from the Nb lines listed in Table~\ref{linelist}. Not all Nb lines which are listed in Table~\ref{linelist} were present in all stars, though, and finding good Nb lines for the intrinsic S stars was even more challenging because of the strong molecular blending. We did not derive the Nb abundance for the intrinsic S star UY Cen as we could not find any well-reproduced Nb line or blend. On average, we used four Nb lines for every extrinsic S star and two for intrinsic S stars. 
These Zr -- Nb abundances will be discussed in Sect.~\ref{ZrNbsec}.

In addition to Zr and Nb, we attempted to determine Sr, Y, Ba, La, Ce, Nd, Sm and Eu abundances in the intrinsic S stars of our sample (Table~\ref{abundtable}).
Finding good Sr lines was impossible for NQ~Pup and V915~Aql, and we could derive the Sr abundance for UY~Cen using only one Sr line. We could find two good Y lines in UY~Cen and even more in NQ~Pup and V915~Aql. The Ba abundance was derived using only one Ba line at 7488.077~\AA\ since other Ba lines were heavily saturated. The La abundance was derived for NQ~Pup and UY~Cen but not for V915~Aql. The Ce and Nd abundances were derived using at least two lines for all three stars. The Sm abundance could be derived only for UY~Cen and NQ~Pup. The Eu abundance could only be obtained for NQ~Pup from one Eu line.
All the atomic lines used for the abundance determination are listed in Table~\ref{linelist} with their excitation potential and $\log gf$.

\subsection{Uncertainties on the abundances} 
\label{uncertainties}
The s-process abundance uncertainties were estimated for the three intrinsic S stars NQ~Pup, V915~Aql and UY~Cen, because they cover an effective temperature range (3300--3700~K) representative of that of S stars, and because the spectra of intrinsic S stars are often more blended than those of extrinsic S stars, so the errors can only be overestimated when applied to our entire sample. Uncertainties are listed in the bottom panels of Tables~\ref{abundtableerr} to \ref{abundtableerr2}. In the upper panels, model~A designates the adopted model, whereas models B--G correspond to models differing by one grid step 
from model~A, each parameter varied at a time.
The abundance resulting from each of these models is then compared to the abundance from model A and these differences are listed as columns $\Delta_{B-A},...,\Delta_{G-A}
$ in  the bottom panels of Tables~\ref{abundtableerr} to \ref{abundtableerr2}.

The stellar parameters of S stars are strongly correlated to one another. Hence, the total error budget should account for this entanglement between the stellar parameters. We followed a method similar to that of \cite{cayrel} to estimate the uncertainties on our abundances due to the entangled stellar parameters. 
After modifying \teff\ by  +100~K or -100~K from model~A, we adjusted the other stellar parameters to obtain an acceptable fit to the global spectrum, as already explained in Sect.~\ref{Sect:atmosph_uncertainties}. We designate this alternate model as 'H' in Tables~\ref{abundtableerr} -- \ref{abundtableerr2}.

\begin{figure*}
    \centering
    \includegraphics[scale=0.45]{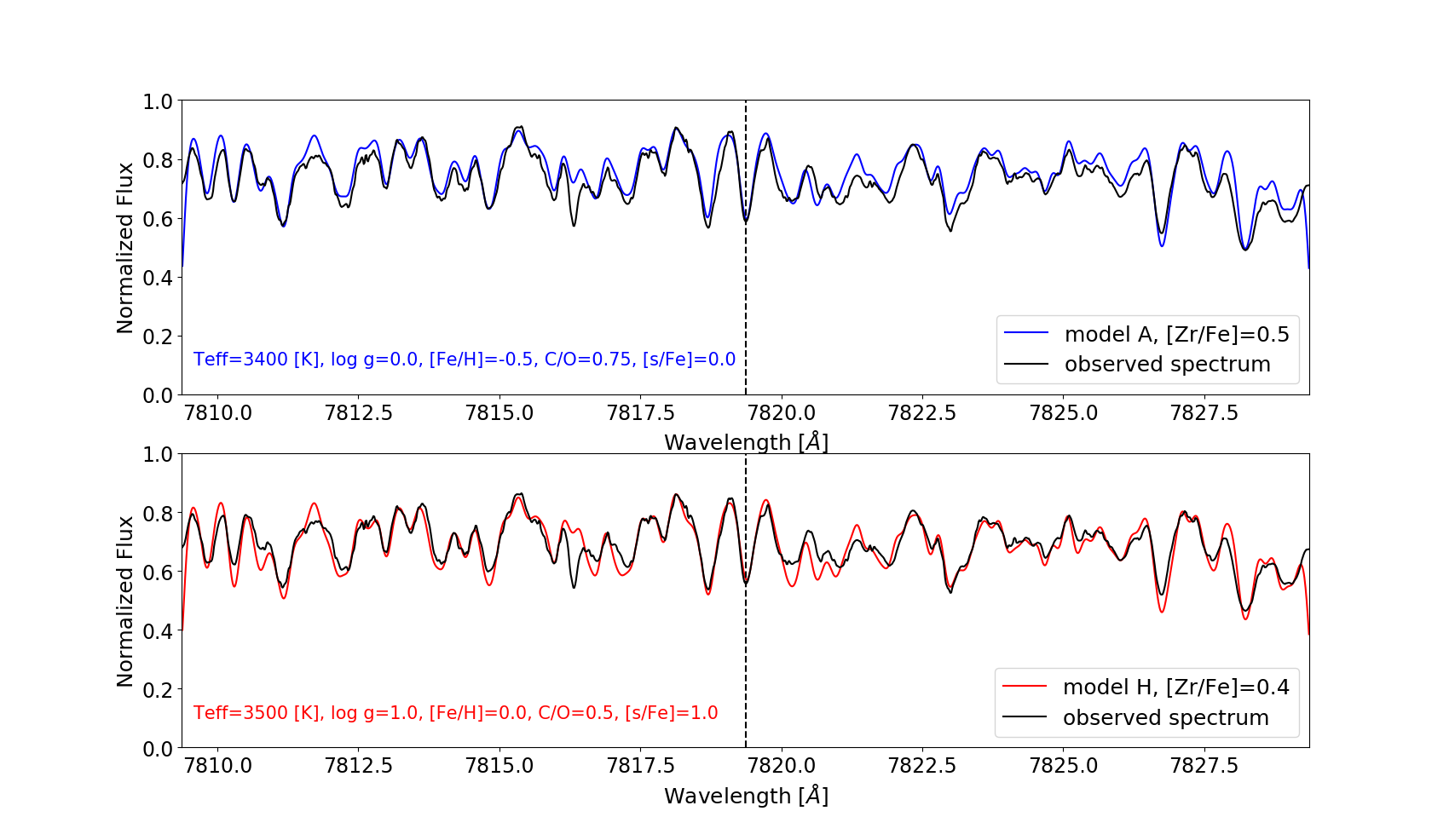}
    \caption{ Comparison between the observed spectrum of V915 Aql (black line) and synthetic spectra corresponding to model~A 
    (top panel, blue line)  and to model~H (bottom panel, red line), in the region surrounding the Zr~I $\lambda$7819~\AA\ line.}
    \label{Fig: normalandE}
\end{figure*}

To further illustrate the strong coupling between atmospheric parameters in S-type stars, Fig.~\ref{Fig: normalandE} compares synthetic spectra around the \ion{Zr}{I} 7819~\AA\ line for models~A and H of V915~Aql. While there are significant differences between the parameters of the two models ($|\Delta T_{\rm eff}|$ = 100~K, $|\Delta \log g| = 1$, $|\Delta $[Fe/H]$| = 0.5$, $|\Delta $C/O$| = 0.25$, $|\Delta $[s/Fe]$| = 1$), 
it is extremely difficult to decide which model has to be preferred from this plot alone. This reveals an intrinsic uncertainty existing for S-star atmospheric parameters. However, our technique consisting in a global assessment combining a $\chi^2$ minimization with gravity  determined from the Gaia parallax
can help to solve this puzzle. Indeed, the $\log g$ value associated with model~H for V915~Aql and UY~Cen is incompatible with the "Gaia $\log g$". Furthermore, for V915 Aql, model H is inconsistent with the derived s-process overabundance: model H has [s/Fe] = +1 dex, whereas the individual abundances do not exceed 0.5 dex. Both arguments support the selection of model~A over model~H.

\begin{figure*}
    \centering
    \includegraphics[scale=0.4]{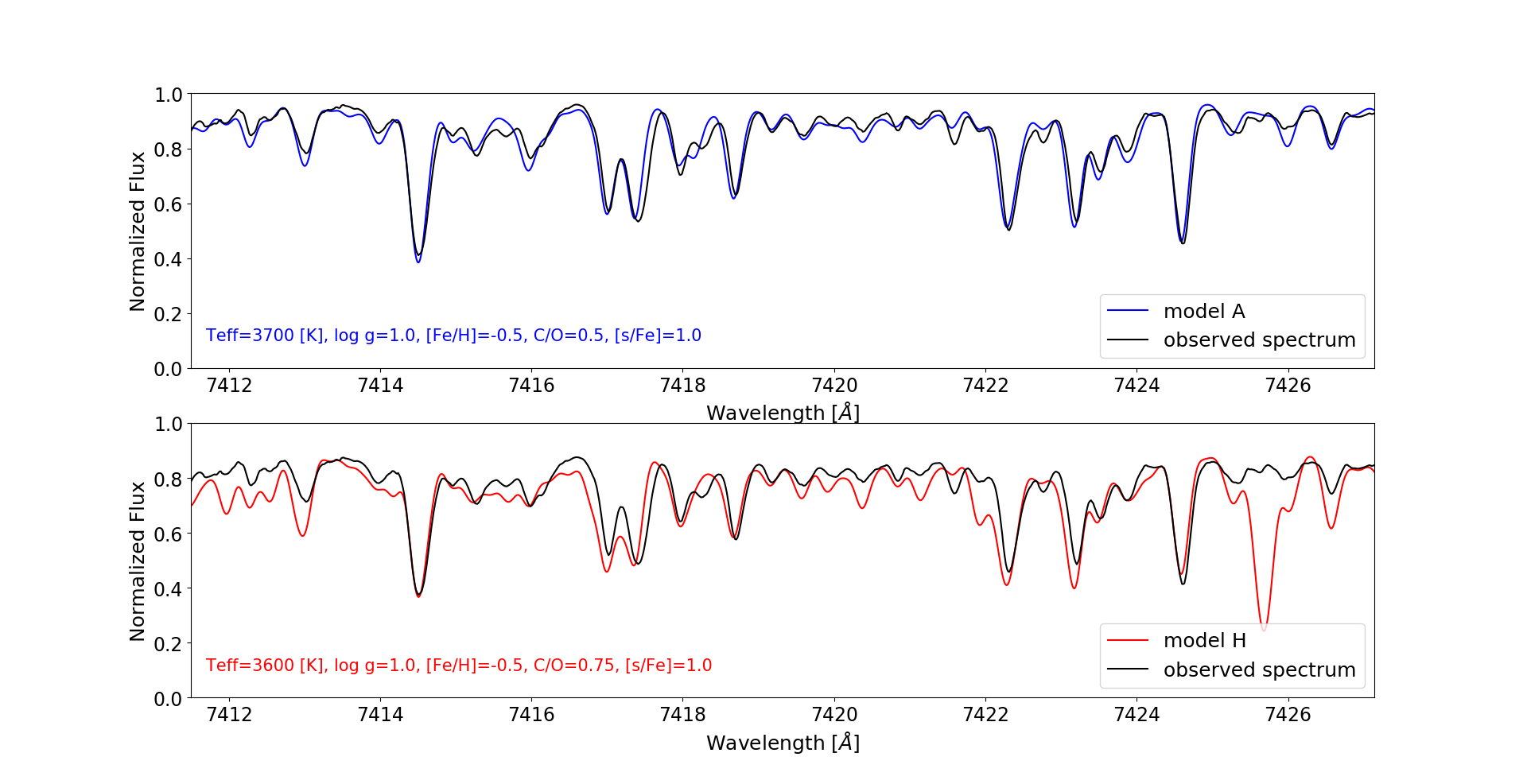}
    \caption{{
    Same as Fig.~\ref{Fig: normalandE} for NQ~Pup in one of the TiO band.}}
    \label{Fig: normalandE_NQP}
\end{figure*}

For NQ~Pup, the situation is more tricky, models A and H 
only differing by $|\Delta T_{\rm eff}|$ = 100~K and $|\Delta $C/O$| = 0.25$.  However, the comparison of the spectral fit in the TiO bands for models A and H clearly favours 
model~A (Fig.~\ref{Fig: normalandE_NQP}).

The expected uncertainty on the abundances due to the changes of multiple stellar-parameter (column $\Delta_{H-A}$ of Tables~\ref{abundtableerr} to \ref{abundtableerr2}) is estimated as the difference between the abundances derived from models H and A (the reference model).

\subsection{Abundance profiles of the three intrinsic S stars NQ~Pup, UY~Cen, and V915~Aql}
\label{Sect: abund-profile}

\begin{table*}
\caption{\label{abundtable} Elemental abundances for NQ~Pup, UY~Cen, and V915~Aql, along with the standard deviation due to line-to-line scatter. The columns entitled $N$ list the number of lines used to derive the abundances. }
\centering
\begin{tabular}{l c c | c c c c | c c c c | c c c c}
\hline
 & & &\multicolumn{4}{c}{NQ Pup}&\multicolumn{4}{c}{UY Cen}& \multicolumn{4}{c}{V915 Aql} \\
\hline
 &$Z$& $\log {\epsilon^a}_\odot$ & $\log \epsilon$ & $N$ & [X/H] &[X/Fe] & $\log \epsilon$&$N$& [X/H] &[X/Fe] & $\log \epsilon$ &$N$&[X/H] &[X/Fe] \\
\hline
C & 6 & 8.43 &8.06 &  & -0.37 & -0.07 & 8.46 & & 0.03 & 0.33 & 8.24 & & -0.19 & 0.31 \\
N & 7 & 7.83 &8.74 &  & 0.91  & 1.21  & 7.80  & & 0.00  & 0.27 & 7.60  & & -0.2 & 0.3 \\
O & 8 & 8.69 &8.36 &  &-0.33  & -0.03 & 8.46  & & -0.23 & 0.07 & 8.36  & &-0.33 & 0.17\\ 
Fe & 26 & 7.50 & 7.20$\pm$  0.07 & 10&-0.30 & &7.18 $\pm$0.15  & 6 & -0.32& & 7.0 $\pm$0.16& 10 & -0.50 &\\
Sr I & 38 & 2.87 & & & & & 4.0$\pm$0.0 & 1 & 1.13 & 1.43 & & & & \\
Y I  & 39 & 2.21 & 2.1$\pm$ 0.18 & 4 & -0.09 & 0.21 & 3.13 $\pm$ 0.41 &3&0.92 & 1.22 & 1.9 $\pm$0.00 &1& -0.31 & 0.19 \\
Y II & 39 & 2.21 & 2.3$\pm$ 0.00 & 1 & 0.09 & 0.39 & & & & & 2.0 $\pm$ 0.00 &1&-0.21&0.29 \\
Zr I & 40 & 2.58 & 2.87$\pm$0.17 & 2 & 0.29 & 0.59 & 3.30 $\pm$0.0  &1&0.72& 1.04 & 2.4 $\pm$ 0.28 & 2& -0.18 & 0.32\\
Nb I & 41 & 1.46 & 1.27$\pm$0.17 &6& -0.19 & 0.11 &  & & &  & 1.00$\pm$0.00 & 1&-0.46 & 0.04\\
Ba I & 56 & 2.18 & 2.35$\pm$0.00 & 1 &0.17 & 0.47 & 2.40$\pm$0.0 &1&0.22 & 0.52 & 2.2$\pm$0.00 &1& 0.02 & 0.52\\
La II & 57 & 1.10 & 1.50$\pm$0.14 & 2& 0.36 & 0.66 & 1.9 $\pm$ 0.0 & 1 & 0.8 &1.1 & & & & \\
Ce II & 58 & 1.70 & 1.50$\pm$0.03 &3 & -0.2 & 0.10 & 1.96$\pm$0.05 &3&0.26 & 0.58 & 1.30 $\pm$0.00&3 & -0.4 & 0.10\\ 
Nd I & 60 & 1.42 & & & & & 2.60 $\pm$ 0.0 & 1 & 1.18 & 1.48 & & & & \\
Nd II & 60 & 1.42 & 1.57$\pm$0.35 & 4& 0.15 & 0.45 & 2.90$\pm$0.0 &1& 1.48 & 1.78 & 1.35 $\pm$ 0.14 &2&-0.07 & 0.43\\
Sm I & 62 & 0.96 & 1.00$\pm$0.00 & 1& 0.04 & 0.34 & && &  & & & & \\
Sm II & 62 & 0.96 & 1.00$\pm$0.00 & 2& 0.04 & 0.34 & 1.00$\pm$0.0 &1&0.04 & 0.34 & & & & \\
Eu II & 63 & 0.52 &0.00$\pm$0.00& 1 &-0.52&-0.22 & & & & & \\
\hline
\end{tabular}
\end{table*}

\begin{figure}      
\begin{centering}
    \mbox{\includegraphics[scale=0.35]{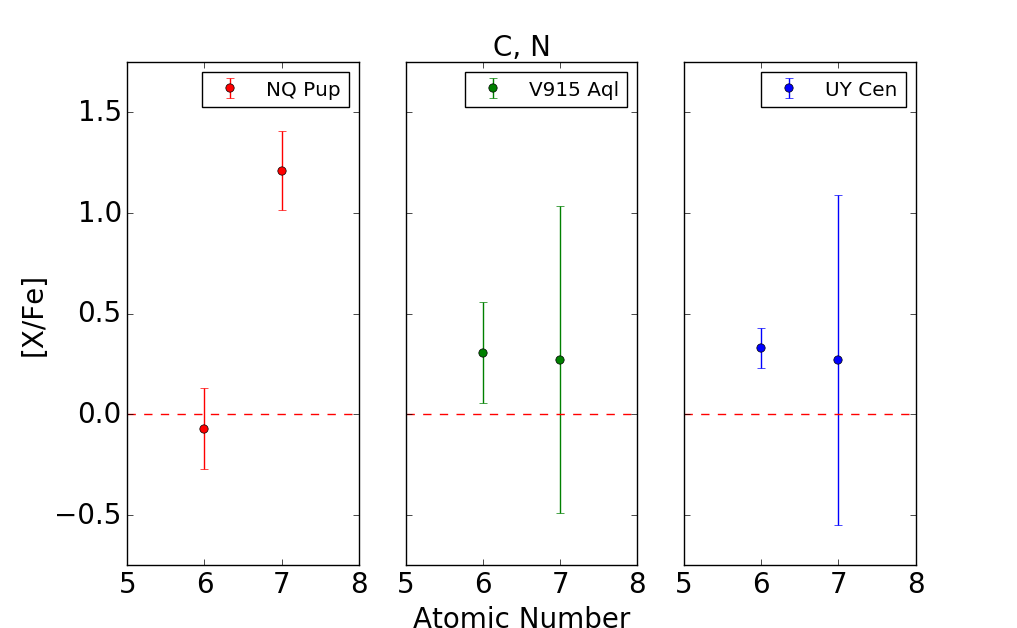}}   
    \mbox{\includegraphics[scale=0.35]{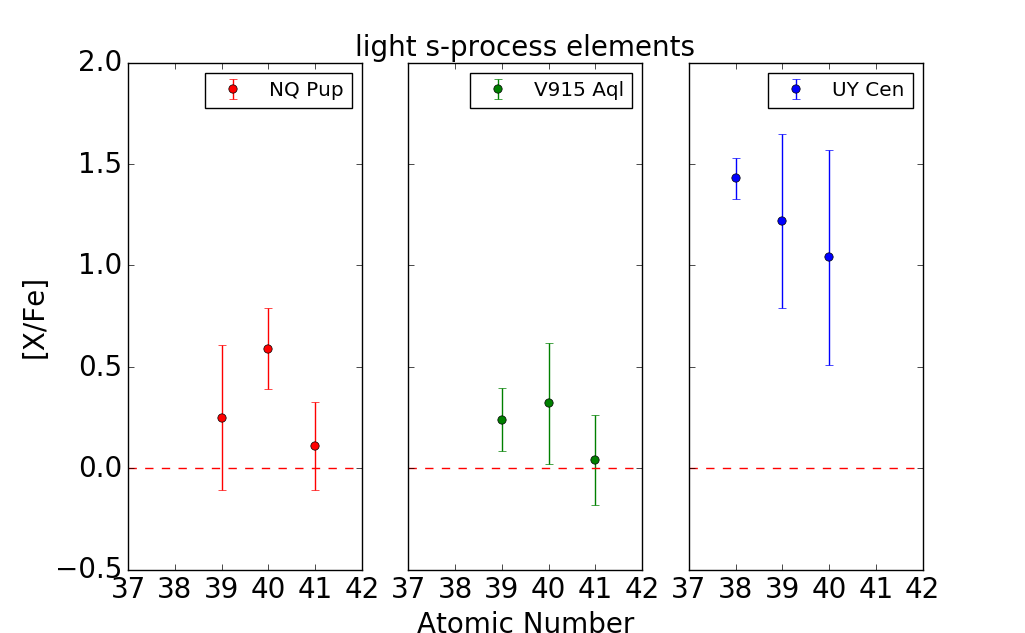}}
    \mbox{\includegraphics[scale=0.35]{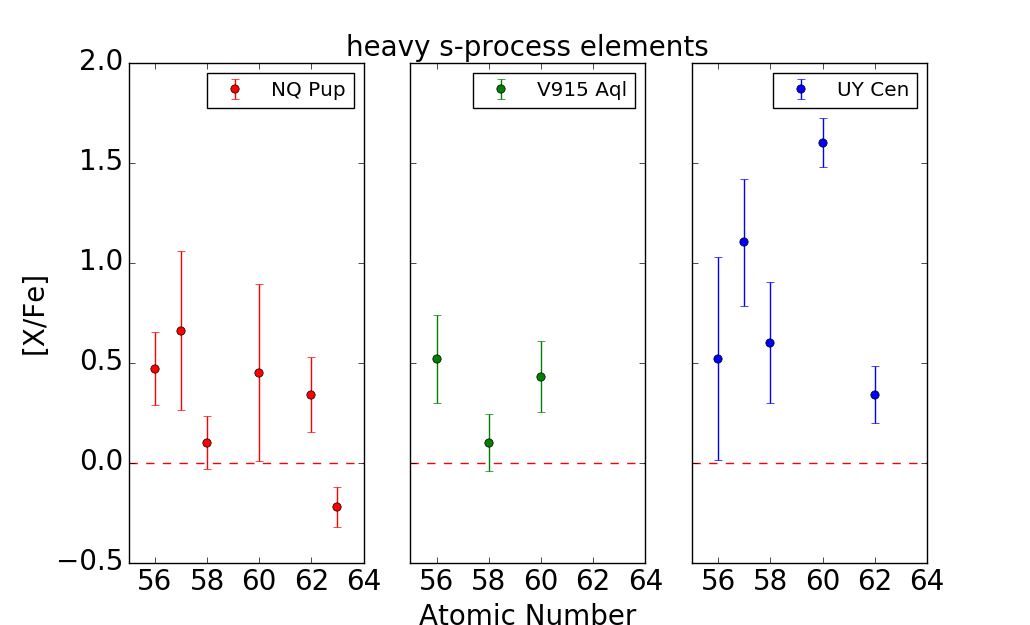}}
    \caption{\label{abundprofile} Abundances of the three intrinsic S stars, with C and N in the top panel, light s-process elements in the middle panel and heavy s-process elements in the bottom panel. The error bars on the elemental abundances represent the total uncertainty calculated by quadratically summing the $\Delta_{H-A}$ values from the last columns of Tables~\ref{abundtableerr} -- \ref{abundtableerr2} with the standard deviation due to line-to-line scatter, plus an extra 0.1 dex as uncertainty due to continuum placement. }
\end{centering}
\end{figure}

As a star evolves on the AGB, it is subject to several episodes of third dredge-up resulting into progressive surface enrichment of carbon and s-process elements. 
The direct monitoring of this evolution of abundances along the AGB 
requires to consider stars of similar masses and metallicities. Unfortunately, as we will show in Sect.~\ref{HRDdiscuss}, the three intrinsic S stars of our sample lie on evolutionary tracks of different masses ($M_{\rm NQ~Pup} \sim 2.5~$M$_{\odot}$, $M_{\rm UY~Cen} \sim 3~$M$_{\odot}$, $M_{\rm V915~Aql} \sim 1~$M$_{\odot}$), even though NQ~Pup and UY~Cen have similar metallicities ([Fe/H]~$\sim -0.3$) close to that of V915 Aql ([Fe/H]~$\sim -0.5$). Therefore the expected correlation between the evolutionary stage and the overabundance level cannot be readily investigated, even though UY~Cen has the largest s-process enrichment and is more evolved on the AGB as compared to NQ~Pup.

We present the  abundances of these three intrinsic S stars in Table~\ref{abundtable}, and discuss them in turn in the remainder of this section.

\medskip
\textit{NQ Pup.}\medskip\\
Our \teff, $\log g$ and [Fe/H] values match well those of \cite{wylie}, but their C/O $> 1$ is larger than our estimate of C/O = 0.5, which is in agreement with that of Neyskens et al. (2015). 
Our metallicity is also very similar to that of  \cite{pieter2015} ([Fe/H]~=~-0.30 and -0.31, respectively). 

The s-process abundance profile of NQ~Pup (left row of Fig.~\ref{abundprofile}) reveals a mild enhancement in s-process elements (and even no detectable Ce enhancement), well in line with the nearly solar C/O.
Our \ion{La}{II}, \ion{Nd}{II}, and \ion{Eu}{II} abundances agree with those from \cite{wylie} within  error bars.
We report lower [Y/Fe] and [Zr/Fe] values as compared to \cite{wylie}. Our Zr abundance differs by 0.17~dex from that of \cite{pieter2015} despite the fact that the same Zr lines have been used in both studies. Our effective temperature differs by 200~K, but the sensitivity of the Zr abundance on \teff\ is too weak to fully account for the observed difference (Table~\ref{abundtableerr}).
The abundance difference is thus probably due to differences in the continuum placement.


\medskip
\textit{V915 Aql.}\medskip\\
 No chemical abundances 
 were reported so far for this Tc-rich S star. Though the carbon content in V915~Aql  (C/O $=0.75$) is slightly higher than in NQ~Pup, we do not measure a stronger s-process enrichment. From the middle panel of Fig.~\ref{abundprofile}, we see that the light and heavy s-process elements reach similar overabundance levels in both stars. We confirm that
 we do observe indications of a moderate s-process overabundance in V915~Aql, including the presence of Tc. Given its low-mass (1~M$_\odot$) inferred from its location in the HR diagram (Fig.~\ref{HRD}), this conclusion is of importance in the framework of stellar evolution, as we will further discuss in Sect.~\ref{Sect:nucleosynthesis}.

 As for NQ~Pup, uncertainties on \teff~have the strongest impact on the abundance determination.

\medskip
\textit{UY Cen.}\medskip\\
Abundance determination in UY Cen is a challenging task because of its low temperature and C/O close to unity. It is consequently an SC star (consistent with its S6/8 spectral type). Abundance analyses of stars of that kind  are known to be challenging  (\citealp{catchpole}, SVE17). The gaps present in the UVES spectral coverage even strengthen  the challenge. 
The elemental abundances were derived from few lines only and to vindicate our results, we compared them with the abundances from \cite{catchpole}, keeping in mind however that SC-like atmospheric models were not available at the time. Catchpole reported that all  elements heavier than Fe are overabundant by 0.6~dex in UY~Cen. We observe  a similar overabundance level (right panel of Fig.~\ref{abundprofile}). Catchpole scaled his abundances with respect to Ti and obtained [Ti/H]~=~-1.05 and [Fe/Ti]~=~0.3 which results in [Fe/H]~=~-0.75. On the contrary, we derive a metallicity [Fe/H]~=~-0.3. This result is obtained from a selection of very good non-blended and non-saturated Fe lines. 

Most s-process elements are highly enriched in UY Cen with [X/Fe]~$ > 1$, in particular Sr, Y, Zr, La, and Nd.
UY Cen is significantly more s-process enriched than NQ Pup (Fig.~\ref{abundprofile}).
This is consistent with the positions of these two stars in the HR diagram where UY~Cen appears more evolved on the TP-AGB (middle panel of Fig. \ref{HRD}), though on an evolutionary track with a higher mass.

\begin{table*}
\caption{\label{abundtableerr} Sensitivity of the elemental abundances of NQ~Pup upon variations of the atmospheric parameters. A dash in the $\Delta$ column indicates that the agreement between the observed and the synthetic spectra was too poor and that the (unique) line usually providing the abundance for the considered element had to be rejected.}
\centering

\begin{center}
\setlength\extrarowheight{-2pt}
\begin{tabular}{c c c r c c c}

\hline
Model    & \teff & $\log g$ & [Fe/H] & C/O & [s/Fe]  &$\chi_{t}$\\
         &  (K)      & (cm s$^{-2}$)         & (dex)  &  & (dex)  &  (km s$^{-1}$)\\ 
\hline
A& 3700&  1.0 &  -0.5&  0.50&  1.00&  2.0 \\
B& 3600&  1.0 &  -0.5&  0.50&  1.00&  2.0 \\
C& 3800& 1.0 &  -0.5&  0.50&  1.00& 2.0 \\
D& 3700&  0.0 &  -0.5&  0.50&  1.00&  2.0\\
E& 3700&  1.0 &  0.0&  0.50&  1.00&  2.0 \\
F& 3700&  1.0 &  -0.5&  0.50&  1.00&  1.5\\
G& 3700&  1.0 &  -0.5&  0.75&  1.00&  2.0 \\
H& 3600& 1.0 &  -0.5&  0.75&  1.00&  2.0 \\
\hline

\end{tabular}
\end{center}

\begin{tabular}{c rrrrrrr}
\hline
Element & $\Delta_{B-A}$  & $\Delta_{C-A}$ & $\Delta_{D-A}$&$\Delta_{E-A}$&$\Delta_{F-A}$&$\Delta_{G-A}$&$\Delta_{H-A}$\\
\hline
[N/Fe]&-0.02&-0.17&-0.72&-0.21&-0.13&-1.05&-0.11\\

[Fe/H]&0.01&-0.24&-0.30&-0.20&-0.04&-0.16&0.00\\

[Y/Fe]&-0.07 &0.22 &0.00 &0.04 &0.03 &-0.04 &-0.30 \\

[Zr/Fe]&0.08&0.37&0.13&0.13&0.22&0.04&-0.02\\

[Nb/Fe]&0.03&0.17&0.13&0.03&0.02&0.11&0.09\\

[Ba/Fe]&-0.26&-0.11&-&-&-&0.01&-0.15\\

[La/Fe]&-0.22&-0.12&0.09&-0.16&-0.32&-0.50&-0.36\\

[Ce/Fe]&-0.01&0.24&-0.15&0.30&0.04&0.11&0.08\\

[Nd/Fe]&-0.24&0.10&-0.12&0.09&0.00&-0.16&-0.25\\

[Sm/Fe]&0.19&0.24&0.30&0.05&0.34&0.12&-0.16\\

[Eu/Fe]&0.17&0.31&0.48&0.38&0.22&0.46&-\\
\hline
\end{tabular}
\end{table*}

\begin{table*}
\caption{\label{abundtableerr3} Same as Table~\ref{abundtableerr} for V915~Aql.}
\centering

\begin{center}
\setlength\extrarowheight{-2pt}
\begin{tabular}{c c c r c c c}
\hline
Model    & \teff & $\log g$ & [Fe/H] & C/O & [s/Fe] &$\chi_{t}$ \\
         &  (K)      & (cm~s$^{-2}$)         & (dex)  &   & (dex)  &  (km~s $^{-1}$)  \\ 
\hline 
A & 3400 & 0.0 & -0.5 & 0.75 & 0.00 &  2.0 \\
B & 3300 & 0.0 & -0.5 & 0.75 & 0.00 &  2.0 \\
C & 3500 & 0.0 & -0.5 & 0.75 & 0.00 &  2.0 \\
D & 3400 & 1.0 & -0.5 & 0.75 & 0.00 &  2.0 \\
E & 3400 & 0.0 & 0.0  & 0.75 & 0.00 &  2.0 \\
F & 3400 & 0.0 & -0.5 & 0.75 & 0.00 &  1.5 \\
G & 3400 & 0.0 & -0.5 & 0.50 & 0.00 &  2.0 \\
H & 3500 & 1.0 & 0.0 & 0.50 &1.00 &  2.0 \\
\hline
\end{tabular}
\end{center}

\begin{tabular}{c rrrrrrr}
\hline
Element & $\Delta_{B-A}$  & $\Delta_{C-A}$ & $\Delta_{D-A}$&$\Delta_{E-A}$&$\Delta_{F-A}$&$\Delta_{G-A}$&$\Delta_{H-A}$\\
\hline
[N/Fe]&0.34&-0.10&1.00&0.03&0.83&0.70&0.63\\

[Fe/H]&0.06&-0.10&-0.20&0.00&0.00&0.50&0.10\\

[Y/Fe]&-0.11 &0.15 &0.15 &0.10 &0.20 &0.10 &0.00 \\

[Zr/Fe]&-0.18&0.20&-0.10&-0.20&-0.30&0.00&0.00\\

[Nb/Fe]&-0.36&0.00&0.00&0.00&0.00&-0.20&0.00\\

[Ba/Fe]&0.04&0.20&-0.20&0.00&0.00&-0.20&-0.10\\

[Ce/Fe]&-0.09&0.13&0.12&0.00&0.20&0.10&0.10\\

[Nd/Fe]&0.39&0.25&-0.05&0.05&0.15&-0.05&-0.20\\
\hline
\end{tabular}

\end{table*}

\begin{table*}
\caption{\label{abundtableerr2} Same as Table~\ref{abundtableerr} for UY Cen.}
\centering

\begin{center}
\setlength\extrarowheight{-2pt}
\begin{tabular}{c c c r c c c}
\hline
Model    & \teff & $\log g$ & [Fe/H] & C/O & [s/Fe] &$\chi_{t}$  \\
         &  (K)      & (cm~s$^{-2}$)         & (dex)  &   & (dex)  &  (km~s $^{-1}$)\\ 
\hline 
A &3300 & 0.0 & -0.5 & 0.99 & 1.00 & 2.0 \\
B &3200 & 0.0 & -0.5 & 0.99 & 1.00 & 2.0 \\
C &3400 & 0.0 & -0.5 & 0.99 & 1.00 &  2.0 \\
D & 3300 & 1.0 & -0.5 & 0.99 & 1.00 & 2.0 \\
E & 3300 & 0.0 & 0.0 & 0.99 & 1.00 & 2.0 \\
F & 3300 &  0.0 &-0.5 & 0.99 & 1.00 &  1.5 \\
G & 3300 & 0.0 & -0.5 & 0.97 & 1.00 & 2.0 \\
H & 3200 & 1.0 & 0.0 & 0.99 & 1.00 &  2.0 \\
\hline
\end{tabular}
\end{center}

\begin{tabular}{c rrrrrrr}
\hline
Element & $\Delta_{B-A}$  & $\Delta_{C-A}$ & $\Delta_{D-A}$&$\Delta_{E-A}$&$\Delta_{F-A}$&$\Delta_{G-A}$&$\Delta_{H-A}$\\
\hline
[N/Fe]&-0.10 &0.53 & 0.40& 0.33&0.28 &0.80 &0.80\\

[Fe/H]&0.12 &-0.28 &0.32 &-0.08 &0.09 &-0.18 &0.40\\

[Sr/Fe]&0.20&- & -&0.30 &- &- &0.00 \\

[Y/Fe]&0.22 &0.67 &0.27 & 0.42&0.26 &0.67 &0.12\\

[Zr/Fe]& -0.22& 0.48&-0.32 &-0.02 &0.11 & 0.08&-0.52\\

[Ba/Fe]&-0.40 &0.70 &-0.20 &-0.10 &0.03 &0.30 &-0.50\\

[La/Fe]&-0.60 &0.10 &-0.50 &-0.40 &-0.27 &0.10 &-0.30\\

[Ce/Fe]&-0.15 &0.25 &0.30 &-0.11 &0.02 &0.14 &0.28\\

[Nd/Fe]&-0.10 &0.60 &0.30 &0.10 &- &0.10 &-0.07\\

[Sm/Fe]&-0.40 &0.00 &-0.20 &-0.40 &-0.37 &-0.40 &-0.10\\
\hline
\end{tabular}

\end{table*}

\section{Comparison of abundances with  nucleosynthesis predictions}
\label{Sect:nucleosynthesis}

\begin{figure*}[ht]      
\begin{centering}
    \includegraphics[scale=0.7]{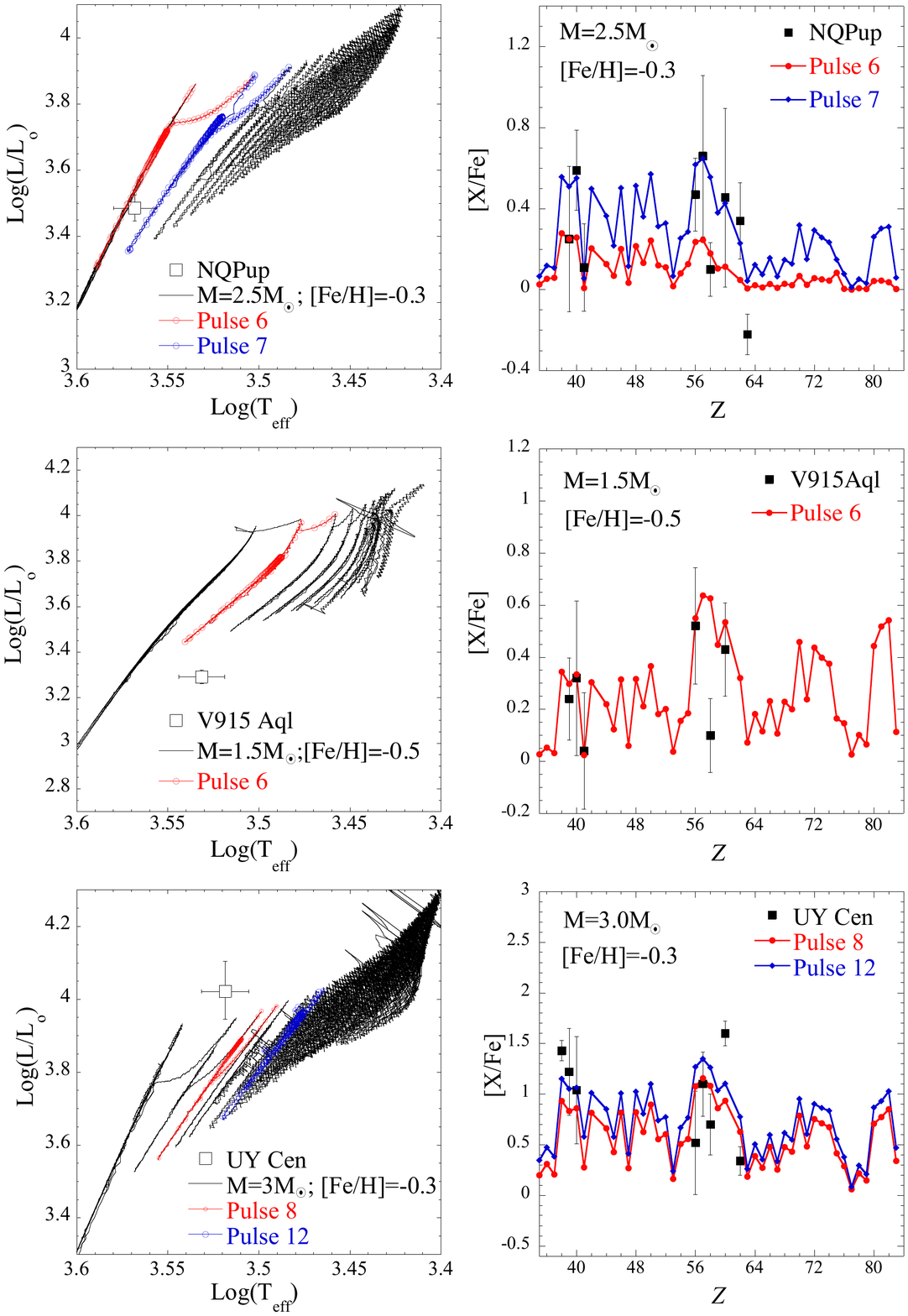}
    \caption{\label{nucleo}
    Left panels: Location of the three intrinsic S stars in the HR diagram, compared with STAREVOL tracks of the corresponding metallicity. Right panels: Predicted abundance distribution of the three intrinsic S stars. Top panel: NQ Pup, central panel: V915 Aql, and bottom panel: UY Cen. }
\end{centering}
\end{figure*}

\begin{figure}
    \centering
    \includegraphics[scale=0.5,trim={1.7cm 0cm 1cm 0cm}]{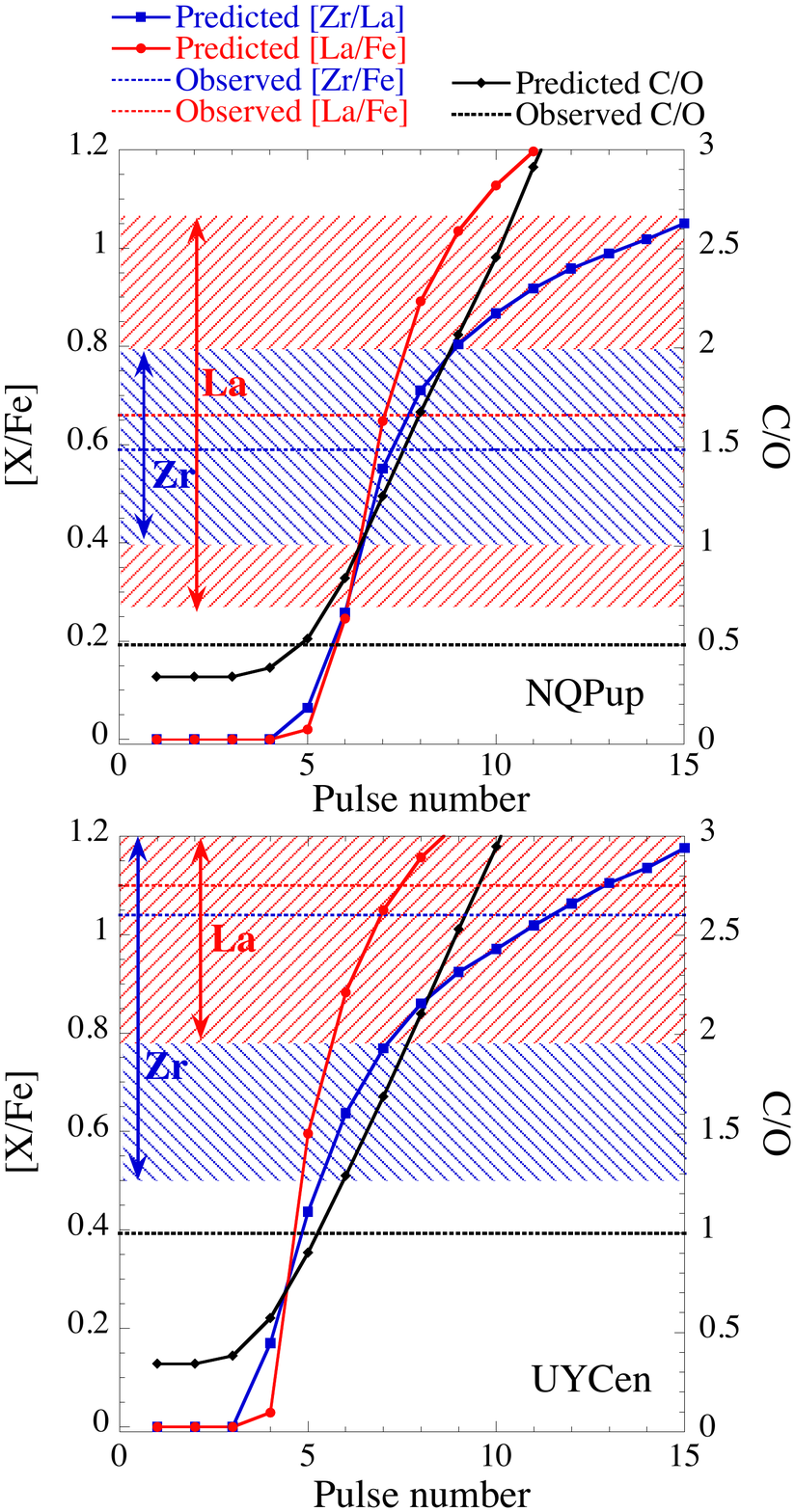}
    \caption{Observed (hatched regions)  La and Zr abundances compared to predicted abundances (red and blue dotted curves) as a function of thermal pulse number. For comparison, the evolution of the C/O is also displayed}
    \label{LaZr}
\end{figure}

\subsection{Stellar models and nucleosynthesis calculations}
\label{sect:starevol}

To locate the stars in the HR diagram, evolutionary tracks have been computed with the
STAREVOL code whose detailed description can be found in  \cite{Siess1} and
\cite{Siess2}. In short, this code considers the radiative opacity tables of
\cite{Iglesias} above 8000~K and of \cite{ferguson} at lower temperatures. Opacity
enhancement due to the formation of molecules in carbon-rich atmospheres is also
accounted for following the formulation of \cite{Marigo2002}. The \cite{schroder}
prescription for the mass-loss rate is considered up to the beginning of the AGB phase
and then the \cite{wood} formulation is applied. To follow the abundances of heavy
elements, we use a nuclear network of 411 species with all the relevant reactions for the
s-process nucleosynthesis. We adopt the solar abundance distribution of
\cite{Asplund-2005} and consider overshooting below the envelope at all times. A
diffusion equation is used to simulate the partial mixing of protons in the C-rich layers
at the time of the third dredge-up. Following the formalism of Eq.~9 of
\cite{goriely&siess}, the diffusive mixing (DM) parameters adopted in our calculations
are $D_{\rm min} =  10^9$~cm$^2/$s and $p =  5$, where $D_{\rm min}$  is the value of
the diffusion coefficient at the  innermost boundary of the diffusive region and $p$  is
an additional free parameter defining the slope of the exponential decrease of the
overshoot diffusion coefficient with depth.

\subsection{The s-process abundance distribution}
\label{Sect:s-process-abun-discussion}

The observed abundance profiles of the 3 intrinsic S stars have been compared with nucleosynthesis predictions from  the STAREVOL code.
The surface abundance distribution displayed in the right panels of Fig.~\ref{nucleo} corresponds to the outcome of s-process nucleosynthesis in an AGB model with the same metallicity as the target star (as listed in Table~\ref{abundtable}), and a mass matching the star location in the HR diagram (left panels of Fig.~\ref{nucleo}). The listed mass and pulse number are indicative. The pulse number is chosen in order to  optimally match both the overabundance level {\it and} the location in the HR diagram. 
 Although such an agreement is not guaranteed {\it a priori},  
it is reached at least for NQ Pup. 
For that star, there is a good match between models  and observations not only for luminosities and temperatures, but also for most abundances (except Ce and Eu). 

The case of V915~Aql is more puzzling (middle panel of Fig.~\ref{nucleo}), since the star lies at an uncomfortably low mass and luminosity for the third dredge-up to occur.   
This object would require a 1~M$_\odot$ model to match its location in the HR diagram, but no s-process surface enrichment is predicted at such a mass because models do not predict the TDU to operate for M~$\lesssim 1.3$~M$_\odot$ \citep[e.g.][ and the present STAREVOL models]{karakas}. To account for V915~Aql surface abundances,  a 1.5~M$_\odot$ model had to be used, but such a model does not agree with the position of V915~Aql in the HR diagram. Nevertheless, abundance predictions for this model reproduce fairly well the overabundance pattern of V915~Aql (except for La).

Finally, concerning UY Cen (bottom panel of Fig.~\ref{nucleo}), 
a 3.5 or 4~M$_\odot$ model would be needed to match its location in the HR diagram. However, such high-mass models do not succeed in matching the rather high level of measured overabundances, because the large envelope mass leads to a strong dilution.
A 3~M$_\odot$ stellar model allows a better match of the abundances, but in turn is too faint to match UY~Cen luminosity in the HR diagram.

\subsection{Comparison between the modelled and measured carbon and s-process enrichments}
\label{Sect:Cs-increase}

The sensitivity of the s-process abundances to the pulse number may be assessed from  Fig.~\ref{LaZr}, which displays the evolution of the Zr and La abundances as a function of the pulse number for the two intrinsic stars with best-matching evolutionary tracks in the HR diagram, namely NQ Pup and UY Cen.
 For NQ Pup, constraints from  Zr and La appear to be mutually compatible with the C/O evolution, pointing at pulse numbers in the range $5-8$. 
For UY Cen however, the measured La and Zr abundances are difficult to reconcile with the model C/O, which increases too rapidly to remain in the range of values characterizing S stars ($0.5<\mathrm{C/O}<1$). 
 Thus, models do not seem to correctly predict the respective rates at which the s-process and carbon abundances increase as a function of pulse number. 
If such a mismatch occurs at a later evolutionary stage along the TP-AGB, it will be far better noticeable as the discrepancy increases with pulse number (compare for example the cases UY~Cen and NQ~Pup).

This discrepancy between s-process and C enrichment levels was already outlined by \cite{goriely-mowlavi-2000}. From their Figs.~12 and 13, it can be inferred that a C/O of unity (which corresponds to the upper limit for an S-type star) implies [Zr/Ti] in the range 0.5 -- 0.8 which is less than the value of [Zr/Fe] = 1.0 estimated for the SC star UY~Cen.

For some objects however, the discrepancy goes in the opposite direction: in the metal-poor ([Fe/H]$=-2.2$) post-AGB star V453~Oph, \citet{Deroo-2005} found a moderate s-process enrichment ([s/Fe]~$\sim 0.5$) not accompanied by a simultaneous C-enrichment. The observed upper limit on the nitrogen abundance in this object excluded a possible CN-cycle destruction of the dredged-up carbon. Moreover, the abundances of elements of similar condensation temperatures as carbon proved that the atmosphere was not depleted in refractory elements like carbon (such a depletion pattern is indeed typical in binary post-AGB stars).
We must thus conclude that a key piece is missing for a successful modelling of the simultaneous s-process and carbon enrichments in AGB and related stars of various metallicities.

A possibly related issue is the marked gap between the (modest) C and s-process enhancements observed in (N-type) carbon stars and those observed in post-AGB stars \citep{2015kenneth}.

\subsection{The Zr -- Nb pair}
\label{ZrNbsec}

\begin{figure}
    \centering
    \includegraphics[scale=0.4]{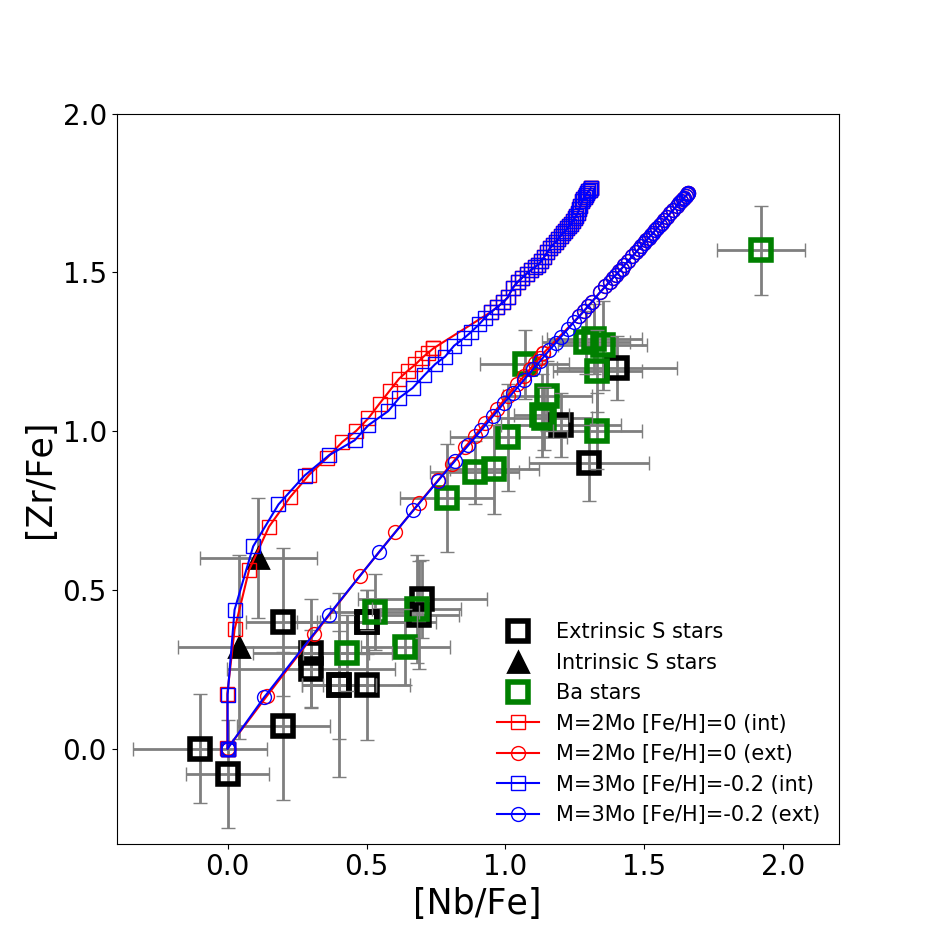}
    \caption{\label{bla}The ([Zr/Fe], [Nb/Fe]) plane. 
    Different kinds of stars are represented by different symbols as labelled in the lower right corner. Abundances for the barium stars are from \protect\citet{drisya}. The error bars on [Zr/Fe] and [Nb/Fe] represent the total uncertainty on the abundances.  
    The red and blue line represent the surface abundance predictions from the models. Each thermal pulse is indicated by a symbol: circles correspond to surface abundances of extrinsic stars, i.e. after the full decay of $^{93}$Zr into Nb, and squares to intrinsic S stars still on the TP-AGB.}
\end{figure}

In extrinsic S stars, the time elapsed since the end of the mass-transfer episode  is sufficient for ${^{93}}$Zr to have fully decayed into mono-isotopic Nb. Actually, mono-isotopic Nb can only be produced by $\beta-$decay of ${^{93}}$Zr (half-life of $1.53\times10^6$~yr). Hence, extrinsic stars should be enhanced in niobium as compared to intrinsic stars.
The Zr~--~Nb pair may therefore be used to confirm the Tc-rich/Tc-poor dichotomy [see the more extensive discussion by \cite{pieter2015} and \cite{drisya}]. Fig.~\ref{bla} shows the trend of [Zr/Fe] vs. [Nb/Fe] for the intrinsic and extrinsic S stars of our sample. 
There is indeed a clear difference between the positions of intrinsic S stars and extrinsic (barium and S) stars in this plane.
The two intrinsic, Tc-rich S stars (NQ~Pup and V915~Aql) are, as expected, Nb-poor, and their Nb and Zr abundances are in excellent agreement with those predicted by AGB nucleosynthesis in 2 or 3~M$_\odot$ stars.

We now compare in this diagram the location of extrinsic S stars with that of barium stars.
Despite their larger error bars (due to the cooler, more blended spectra of S stars), the Nb and Zr abundances of extrinsic S stars are well in line with those of barium stars, as expected since extrinsic S stars are just the cooler analogs of barium stars (see Fig.~\ref{HRDBastars} below). The fact that Nb and Zr abundances of most barium stars in Fig.~\ref{bla} are larger than those of extrinsic S stars is probably due to a selection bias, since Karinkuzhi et al. (2018) deliberately selected highly-enriched barium stars (so that their primordial Nb and Zr abundances could be neglected).

\section{The HR diagram of S stars}\label{HRDdiscuss}

\begin{figure*}      
\begin{centering}
    \mbox{\includegraphics[scale=0.42,trim={1.5cm 0cm 0cm 0cm}]{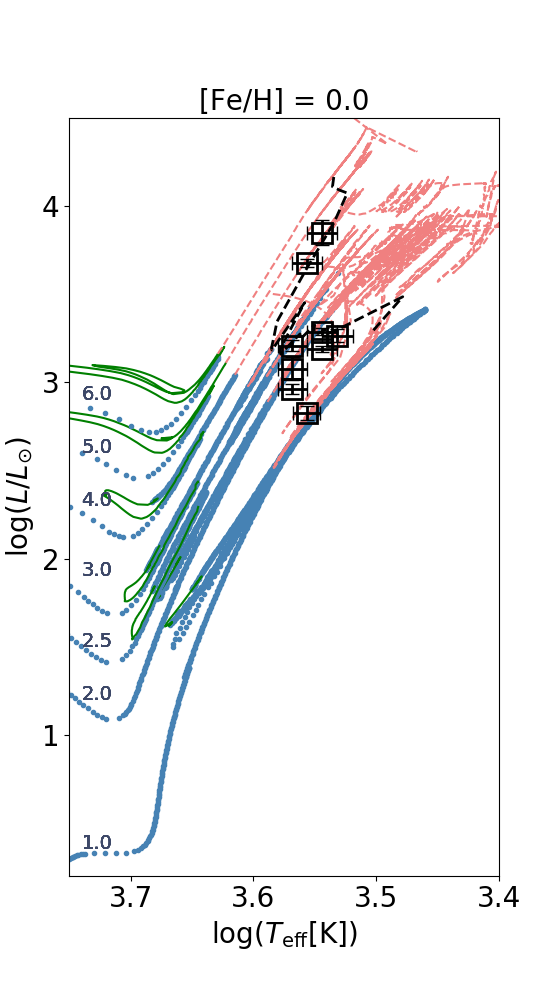}}   
    \hspace{0px}
    \begin{overpic}[scale=0.42,trim={1.3cm 0cm 0cm 0cm}]{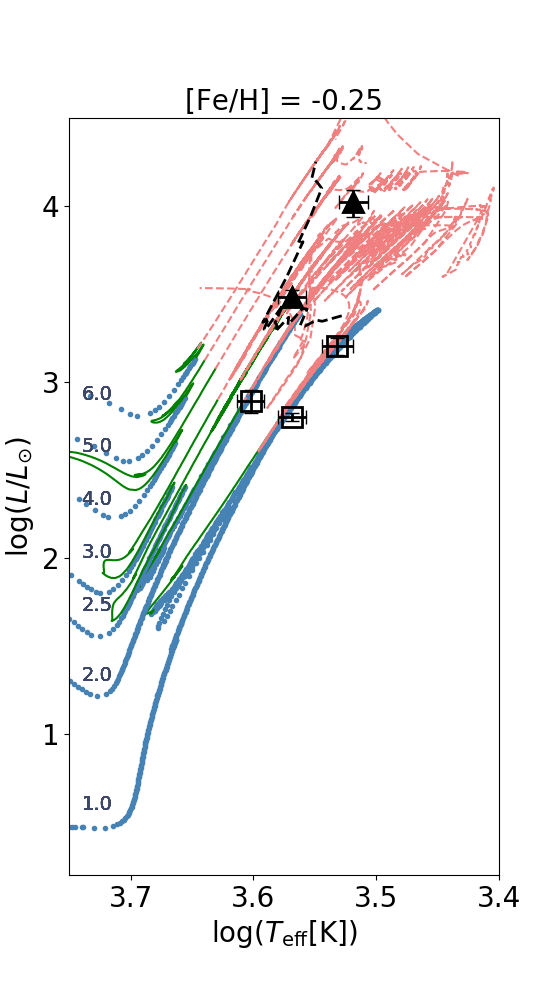}  
     \put(34,78){\textbf{\small{UY Cen}}}
     \put(11,68){\textbf{\small{NQ Pup}}}
    \end{overpic}
    \hspace{0px}
    \begin{overpic}[scale=0.42,trim={1.3cm 0cm 0cm 0cm}]{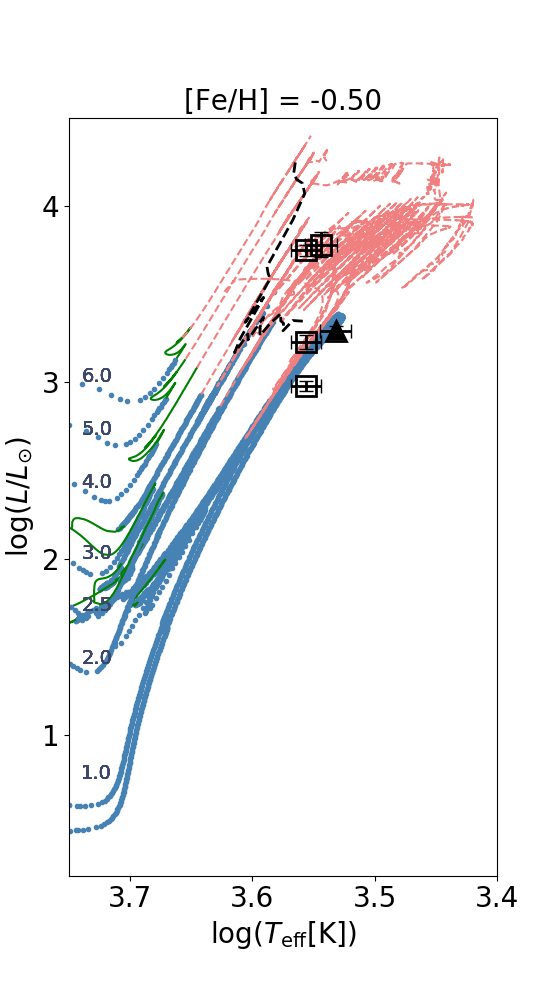}  
     \put(32,65){\textbf{\small{V915 Aql}}}
    \end{overpic}
    \caption{\label{HRD}HR diagram of intrinsic (filled triangles) and extrinsic (open squares) S stars along with the STAREVOL evolutionary tracks corresponding to the closest grid metallicity. 
    The red giant branch is represented by the blue dots, the core He-burning phase by the green solid line, whereas the red dashed line corresponds to the AGB tracks. HD~189581 has been moved by -100~K to avoid overlap with the extrinsic S stars AB~Col and V530~Lyr in the left panel. 
    }
    \label{materialflowChart}
\end{centering}
\end{figure*}

Fig.~\ref{HRD} presents the HR diagram of our sample of S stars using the final stellar parameters from Table~\ref{finalparams} and the stellar evolutionary tracks of the corresponding metallicity computed with the STAREVOL code. The asymmetric error on luminosities was derived by propagating the symmetric error on the parallax. 

The derivation of stellar masses is possible by locating individual stars on evolutionary tracks of the corresponding metallicity. Both metallicity and mass are listed in Table~\ref{finalparams}. Although no uncertainty on the mass is given in Table~\ref{finalparams}, it may be estimated from Fig.~\ref{HRD} by noting that the main sources of uncertainty are the $\pm100$~K uncertainty on the \teff\ derivation, along with 
an  intrinsic error of about $100-150$~K on the location of the evolutionary tracks themselves \citep{Cassisi-2017}.

The two intrinsic stars NQ~Pup and UY~Cen are both located on the TP-AGB, as expected. UY~Cen is located well above the predicted onset of TDU (black dotted line in the middle panel of Fig.~\ref{HRD}, corresponding to the lowest stellar luminosity following the first occurrence of a TDU episode), which is consistent with its large overabundance level in s-process elements. On the contrary, NQ~Pup lies just above the TDU threshold, as already reported by \cite{hipp}. 

The intrinsic (Tc-rich) S star V915~Aql, with \mbox{[Fe/H]~=~-0.5} and located along a 1~M$_{\odot}$ track, constitutes a most puzzling case 
(as already mentioned in Sect.~\ref{Sect:s-process-abun-discussion})
because the third dredge-up is expected to occur only for masses larger than 1.3~M$_\odot$ according to stellar evolution predictions \citep[e.g.][and the present STAREVOL models]{karakas}. V915~Aql is not the only low-mass AGB star where evidence of TDU is observed. \cite{2015kenneth} reached a similar conclusion in their analysis of low-luminosity s-process-rich post-AGB stars.

On the contrary, the HRD location of most extrinsic S stars is compatible with them being on the RGB, confirming the conclusions drawn for S stars from the HIPPARCOS data \citep{hipp}. However, the extrinsic S stars HD~150922, HD~191226, FX~CMa,  and BD~-10$^\circ$1977 (which are the high-luminosity stars along the [Fe/H]~=~0.00 and -0.5 tracks in Fig.~\ref{HRD}) have positions compatible with the early-AGB phase. Interestingly, from the positions of our extrinsic S stars in the HRD, we can infer that the Tc-poor S stars of our sample lie on the early-AGB if $M \ge 2$~M$_\odot$ and on the upper part of the RGB or on the early-AGB if $M \le 2$~M$_\odot$.

\begin{figure}
\begin{center}
\includegraphics[scale=0.27,trim={1cm 0cm 0cm 0cm}]{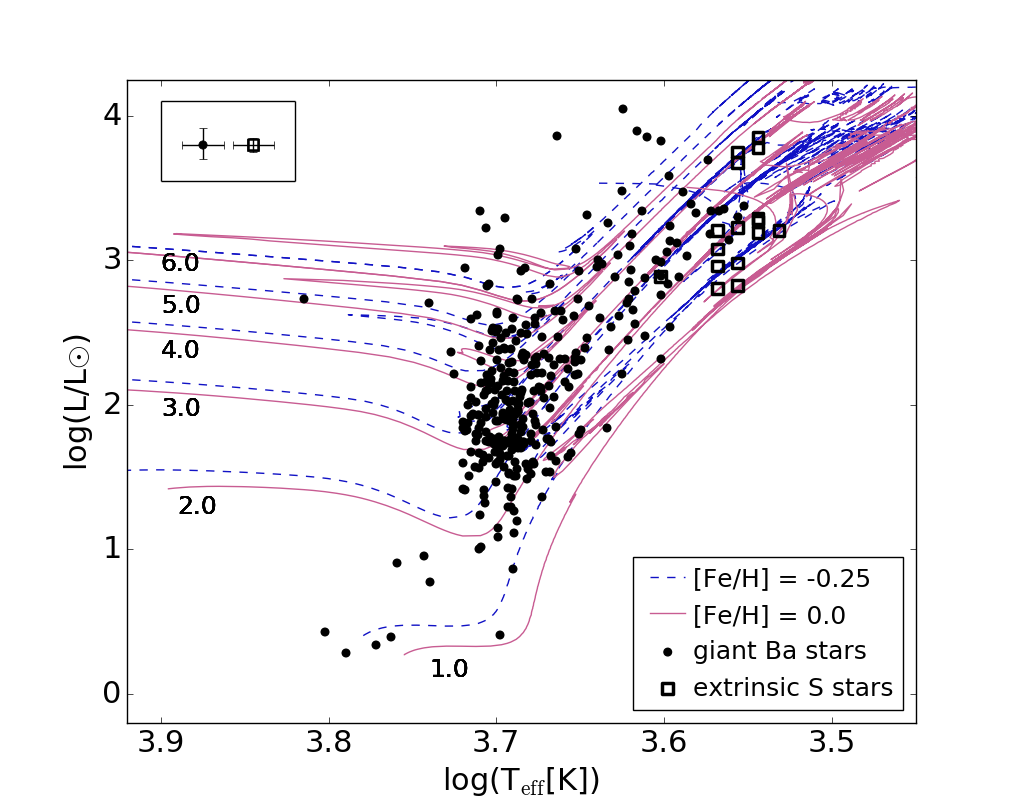}
\caption{\label{HRDBastars} HR diagram of Ba giants and extrinsic S stars superimposed on the STAREVOL evolutionary tracks with [Fe/H] = 0.00 and -0.25. The upper left labels display the typical error bars.}
\end{center}
\end{figure}

\begin{figure}
\begin{center}
\includegraphics[scale=0.5]{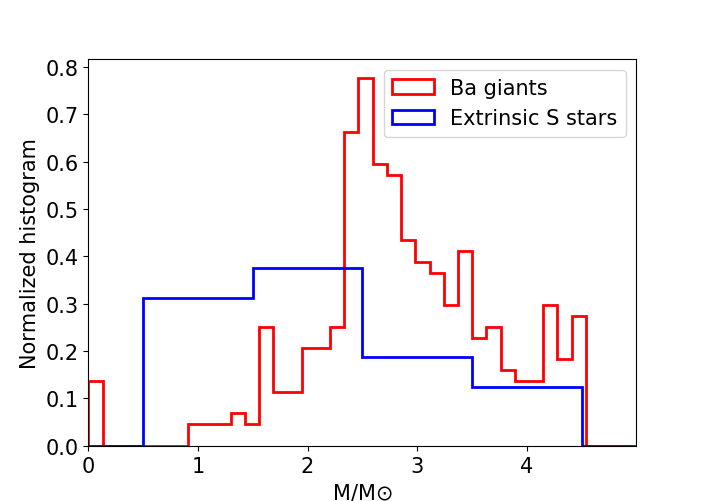}
\caption{\label{histBaS} Histogram of the mass distributions for extrinsic S stars from our sample (listed in Table~\ref{finalparams}) and for the barium giants from \cite{anaBa}.}
\end{center}
\end{figure}

Since extrinsic S stars are believed to be the cooler analogs of barium stars \citep[e.g.][]{jorissen1988}, it is interesting to compare the location of these two families in the HR diagram. We thus added our sample of extrinsic S stars to the barium-star HR diagram of \citet[][their Fig.~7]{anaBa}, resulting in Fig.~\ref{HRDBastars}. These authors used STAREVOL evolutionary tracks of metallicity [Fe/H]=-0.25 while the average metallicity of our extrinsic S-star sample is -0.1. Hence, we also added the solar-metallicity STAREVOL tracks to the HR diagram of Fig.~\ref{HRDBastars}.
As expected, Tc-poor S stars appear as the cooler analogs of barium stars. The reason is that the ZrO molecular bands distinctive of S-type stars only appear below $T_{\rm eff} \approx 4000$~K. The coolest barium stars of \citet{anaBa} have 
$T_{\rm eff} \approx 3600$~K, while our warmest S star has 
$T_{\rm eff}= 4000$~K. The stars located in the transition region between barium and S stars indeed have an hybrid spectral classification. A good example thereof is IT~Vir (HD~121447), classified as SC2 \citep{Ake-1979}, K4Ba \citep{Abia-1998}, K7Ba, and S0.

Another interesting feature emerging from Fig.~\ref{HRDBastars} is that, despite S stars being cooler than barium stars, they  are not necessarily brighter, and this is due to the fact that some barium stars are much more massive than extrinsic S stars.
In fact, the comparison of barium and extrinsic S-star masses  (Fig.~\ref{histBaS}) reveals  that, despite the small number statistics (only 16 extrinsic S stars), S-star masses peak around $1-2$~M$_\odot$, whereas barium-star masses peak around 2.5~M$_\odot$.
This is not surprising, since the RGB extends to cool enough temperatures for the ZrO bands to form (and the star to be classified as type S) only for low-mass stars
(3000~K for 1~M$_\odot$, 3550~K for 2~M$_\odot$), whereas for masses larger than 2.5~M$_\odot$, the RGB tip is at T$_{\rm eff}> 4260$~K.
Hence, barium stars with masses in excess of $\sim 2.5$~M$_\odot$ can only turn into extrinsic S stars on the E-AGB, but those are short-lived, and thus rare.

\section{Infrared excess of the intrinsic S stars}\label{IR}

\begin{figure*}
\begin{centering}
    \mbox{\includegraphics[width=6cm, height=5.5cm, trim={1cm 0.5cm 1cm 1cm}]{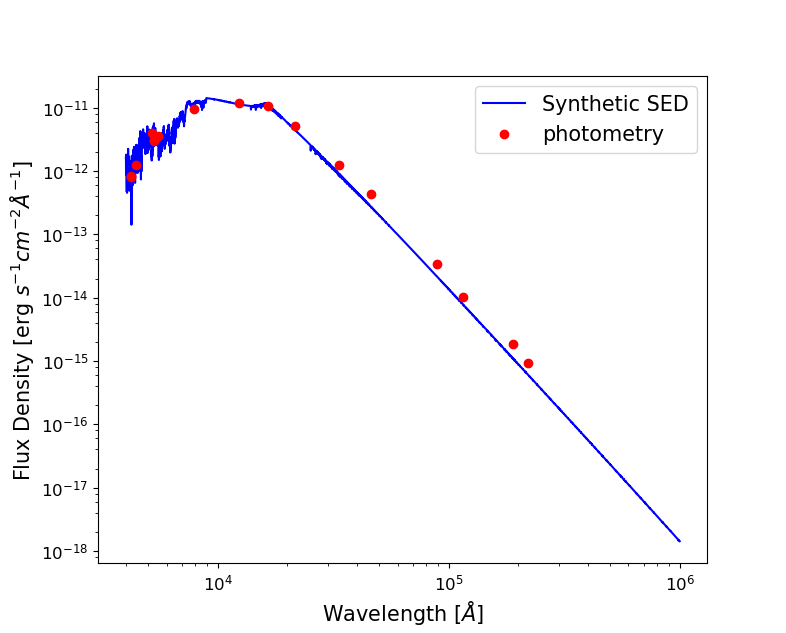}} 
    \hspace{0px}
    \mbox{\includegraphics[width=6cm, height=5.5cm,trim={1cm 0.5cm 1cm 1cm}]{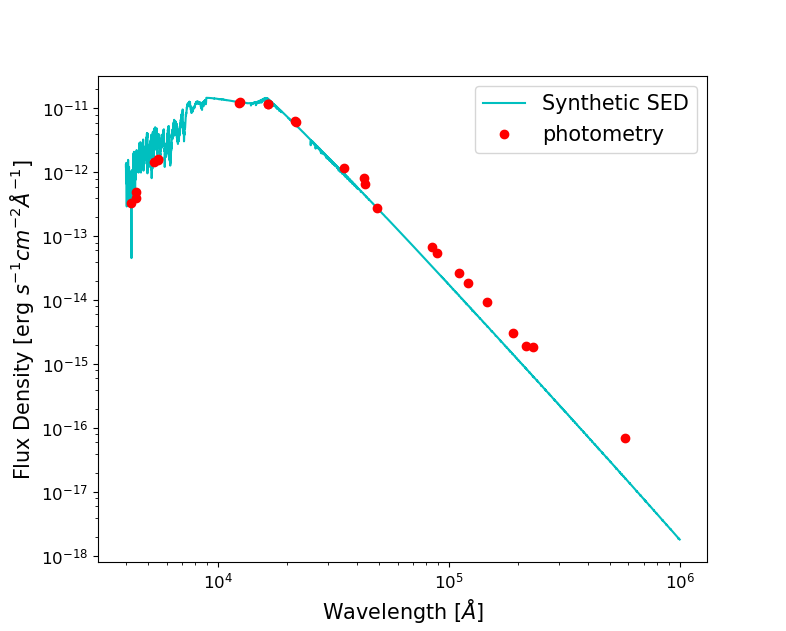}}
    \hspace{0px}
    \mbox{\includegraphics[width=6cm, height=5.5cm,trim={1cm 0.5cm 1cm 1cm}]{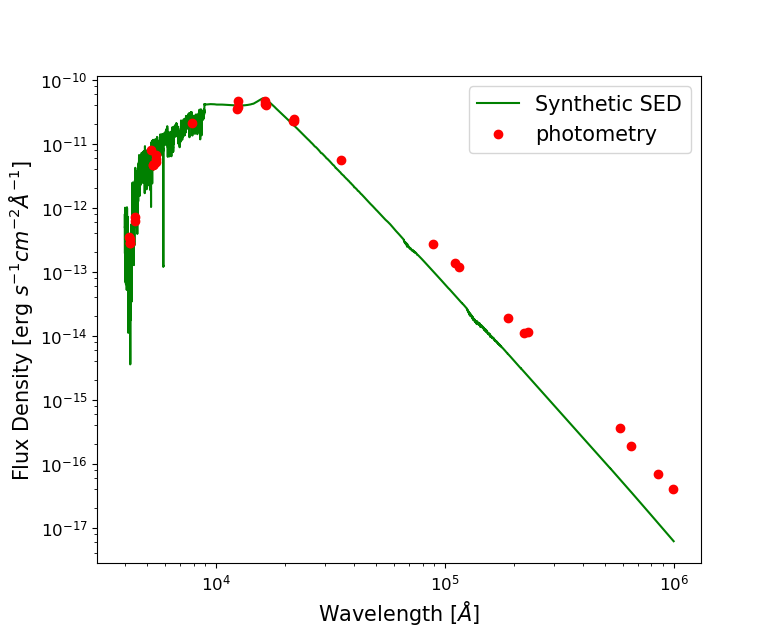}}
    \caption{\label{SEDs} The spectral energy distributions of NQ~Pup, V915~Aql and UY~Cen (from left to right), compared with the best-matching spectroscopic MARCS model.}
\end{centering}
\end{figure*}

Infrared excesses observed for AGB stars are associated with the formation of dust; indeed, mass loss and pulsations bring dust far enough from the warm photosphere to allow dust to condensate.
Fig.~\ref{SEDs} compares the SEDs of the three intrinsic S stars from our sample. 
There is a clear increase of the infrared excess with decreasing effective temperature  for the three stars NQ~Pup, V915~Aql and 
UY~Cen (3700~K, 3400~K, and 3300~K, respectively).
Furthermore, a comparison with Fig.~\ref{abundprofile} shows that this trend is correlated with the s-process enrichment, since UY~Cen is the object with the coolest effective temperature, the largest infrared excess and the largest s-process overabundance.
This trend is also well correlated with the C/O  (C/O=0.50, 0.75 and 0.999 for NQ~Pup, V915~Aql and UY~Cen, respectively; Table~\ref{finalparams}).
These evolutionary indicators point at NQ~Pup and V915~Aql as being relatively unevolved intrinsic S stars (at the beginning of the TP-AGB) while UY~Cen is among the most evolved intrinsic S stars, consistent with both its location in the HR diagram (well above the predicted onset of TDU) and its spectral classification [S6/8 (Table~\ref{basicdata}) or C abundance \citep{LloydEvans-2010}].

\section{Conclusions}

In this paper, we illustrated the specific difficulties faced when trying to derive the atmospheric parameters of S-type stars.
Fitting their SED  for instance 
leads to 
effective temperatures accurate to within 200~K, but 
does not allow one to fix the other stellar parameters ($\log g$, [Fe/H], C/O, [s/Fe]) without ambiguities. We therefore developed a spectral-fitting method making use of the grid of MARCS model atmospheres for S-type stars. This model grid covers the required range in $T_{\rm eff}$, surface gravity, metallicity, C/O, and [s/Fe].
 The spectral-fitting method finds, among the grid of synthetic spectra, those matching best an observed high-resolution, high S/N spectrum. Combined with Gaia DR2 parallaxes and stellar evolution models, it finally allows one to select the most appropriate model, producing  a good agreement at high resolution between synthetic and observed spectra. This provides the necessary basis for deriving reliable surface abundances.
 The Zr -- Nb abundances of extrinsic and intrinsic S stars were especially investigated, and appear to be in good agreement with the  predictions of s-process nucleosynthesis in AGB stars. In particular, a  segregation is observed between intrinsic (Tc-rich) S stars and extrinsic (barium and S) stars, Tc-rich S stars being, as expected, Nb-poor.

The combination of their atmospheric parameters  with their  Gaia DR2 parallaxes also provide the location of S stars in the HR diagram. Extrinsic S stars lie either on the upper part of the RGB (for most of them)  or on the early-AGB, whereas intrinsic S stars like NQ~Pup and UY~Cen are located on the TP-AGB (with UY~Cen lying  well above the onset of TDU episodes). The situation is less clear for the low-mass, Tc-rich S star V915~Aql, pointing at the possible occurrence of TDU episodes in low-mass ($M \sim 1$~M$_\odot$) stars. 
For our intrinsic stars, we find a good qualitative correlation between increasing infrared excesses, increasing C/O, increasing s-process overabundance levels, and decreasing effective temperatures.
However, another puzzle comes from the tension between the measured and predicted rates at which C/O and s-process abundances increase. 

In a forthcoming paper, the Gaia DR2 parallaxes of a larger number of intrinsic S stars will be used in order to constrain the luminosity of the first occurrence of the third dredge-up, 
as well as the luminosity where the C/O reaches unity (i.e. when S stars turn into SC stars).

Extrinsic S stars appear as the cooler analogs of barium stars in the HR diagram. A comparison of their respective masses reveal, however, that barium stars are on average more massive than extrinsic S stars. This results simply from the fact that, to develop their distinctive ZrO bands, S-type stars should be cooler than $\sim4200$~K, and this is possible only close to the RGB tip of low-mass stars. The E-AGB also fulfills this condition, but the shorter time scale associated with that phase does not favour the detection of extrinsic S stars at that stage.

\begin{acknowledgements}
      This research has been funded by the Belgian Science Policy Office under contract BR/143/A2/STARLAB. 
      S.V.E. thanks {\it Fondation ULB} for its support.
      Based on observations obtained with the HERMES spectrograph, which is supported by the Research Foundation - Flanders (FWO), Belgium, the Research Council of KU Leuven, Belgium, the \textit{Fonds National de la Recherche Scientifique} (F.R.S.-FNRS), Belgium, the Royal Observatory of Belgium, the \textit{Observatoire de Gen\`eve}, Switzerland and the \textit{Th\"{u}ringer Landessternwarte Tautenburg}, Germany. This work has made use of data from the European Space Agency (ESA)  mission
      Gaia \url{(https://www.cosmos.esa.int/gaia)},  processed  by the Gaia Data  Processing  and  Analysis  Consortium \url{(DPAC, https://www.cosmos.esa.int/web/gaia/dpac/consortium)}. Funding for the DPAC has
      been provided by national institutions, in particular the institutions participating in the Gaia
      Multilateral Agreement. This research has also made use of the SIMBAD database, operated at CDS, Strasbourg, France. LS \& SG are senior FNRS research associates.
\end{acknowledgements}



\bibliographystyle{aa}
\bibliography{biblio}
\begin{appendix}
\section{Line list}
Table~\ref{linelist} lists the lines used for the abundance analysis.

\begin{table*}
\setlength\extrarowheight{-3pt}
\centering
\caption{Atomic lines used in this study. The last column lists the stars where the corresponding line was used for abundance determination (N: NQ~Pup; V: V915~Aql; U: UY~Cen).}
\label{linelist}
\begin{tabular}{c c c c c c}

\hline
Species& $\lambda$ [\AA] & $\chi$ [eV] & $\log gf$ & Reference & Star\\
\hline\\
Fe I & 7389.398 & 4.301 & -0.460 & \cite{K07} & U\\
 & 7418.667 & 4.143 & -1.376 & \cite{BWL} & V\\
 & 7443.022 & 4.186 & -1.820 & \cite{MFW} & N\\
 & 7461.263 & 5.507 & -3.059 & \cite{K07} & NU\\
 & 7498.530 & 4.143 & -2.250 & \cite{MFW} & VU\\
 & 7540.430 & 2.727 & -3.850 & \cite{MFW} & N\\
 & 7568.899 & 4.283 & -0.773 & \cite{K07} & U \\
 & 7586.018 & 4.313 & -0.458 & \cite{K07} & NV\\
 & 7832.196 & 4.435 & 0.111 & \cite{K07} & U \\
 & 7937.139 & 4.313 & 0.225 & \cite{K07} & U\\
 & 8108.320 & 2.728 & -3.898 & \cite{K07} & NV\\
 & 8698.706 & 2.990 & -3.452 & \cite{K07} & NV\\
 & 8699.454 & 4.955 & -0.380 & \cite{NS} & NV\\
 & 8710.404 & 5.742 & -5.156 & \cite{K07} & NV\\
 & 8729.144 & 3.415 & -2.871 & \cite{K07} & NV\\
 & 8747.425 & 3.018 & -3.176 & \cite{K07} & V\\
 & 8763.966 & 4.652 & -0.146 & \cite{NS} & V\\
Fe II & 7454.035 & 10.562 & -4.130 & \cite{RU} & N\\
Sr I & 4962.259 & 1.847 & 0.200 & \cite{GC} & U\\
Y I & 5630.130 & 1.356 & 0.211 & \cite{MC} & U \\
 & 6402.006 & 0.066 & -1.849 & \cite{K07} & N\\
 & 6435.004 & 0.066 & -0.820 & \cite{HL} & N\\
 & 6557.371 & 0.000 & -2.290 & \cite{K07} & N\\
 & 6793.703 & 0.066 & -1.601 & \cite{K07} & N\\
 & 7802.485 & 1.900 & -1.880 & \cite{CB} &U\\
 & 7812.13 & 5.849 & -3.155 & \cite{K07} & U \\
 & 8800.588 & 0.000 & -2.240 & \cite{CB} & V\\
Y II & 7881.881 & 1.839 & -0.570 & \cite{Nil} & NV \\
 Zr I & 7819.374 & 1.822 & -0.380 & \cite{Zrlines} & NVU\\
 & 7849.365 & 0.687 & -1.300 & \cite{Zrlines} & NV\\
Nb I & 4116.888 & 0.000 & -1.180 & \cite{DLa} & N\\
 & 4195.089 & 0.020 & -0.910 & \cite{DLa} & N\\
 & 4262.053 & 0.130 & -0.560 & \cite{DLa} & N\\
 & 4345.308 & 0.000 & -1.360 & \cite{DLa} & N\\
 & 5189.186 & 0.130 & -1.394 & \cite{DLa} &N\\
 & 5271.524 & 0.142 & -1.240 & \cite{DLa} & V\\
 & 5350.722 & 0.267 & -0.862 & \cite{DLa} & N\\
Ba I & 7488.077 & 1.190 & -0.230 & \cite{MW} & NVU\\
La I & 7068.387 & 0.131 & -1.780 & \cite{CB} & U\\
La II & 4322.464 & 0.173 & -2.377 & \cite{BP} & N\\
 & 5114.559 & 0.235 & -1.030 & \cite{LBS} & N\\
Ce II & 8404.133 & 0.704 & -1.670 & & VU\\ 
 & 8716.659 & 0.122 & -1.980 & \cite{MC} & NVU\\
 & 8772.135 & 0.357 & -1.260 & \cite{PQWB} & NVU\\
Nd II & 4715.586 & 0.205 & -0.900 & \cite{HLSC} &N\\
 & 5276.869 & 0.859 & -0.440 & \cite{MC} & NVU\\
 & 5293.160 & 0.823 & 0.100 & \cite{HLSC} & NV\\
 & 5291.656 & 0.850 & 0.400 & \cite{MC} & U \\
 & 5319.810 & 0.550 & -0.140 & \cite{HLSC} & N\\
Sm I & 4523.178 & 0.101 & -0.839 & \cite{PK} &N\\
Sm II & 4244.696 & 0.277 & -0.810 & \cite{LD-HS} &N \\
 & 4318.926 & 0.277 & -0.250 & \cite{LD-HS} & N\\
 & 8048.681 & 1.746 & -0.370 &  \cite{MC} & U\\
Eu II & |4129.69 & 0.000 & -0.977 & \cite{BP} &N\\
 
\hline

\end{tabular}

\end{table*}

\end{appendix}

\end{document}